\documentclass[iop]{emulateapj}
\usepackage{color}
\usepackage{graphicx}
\usepackage{longtable}
\usepackage{lineno}
\usepackage{draftcopy}
\usepackage{threeparttable}
\usepackage{multirow}
\slugcomment{version \today: fm}
\shorttitle{}
\shortauthors{}

\begin{document}
\title{Low-Mass X-ray Binaries and Globular Clusters\\ 
	Streamers and Arcs in NGC4278.}
\author{R.~D'Abrusco\altaffilmark{1}, G.~Fabbiano\altaffilmark{1} \& N.J.~Brassington\altaffilmark{2}}

\altaffiltext{1}{Harvard-Smithsonian Astrophysical Observatory, 60 Garden Street, Cambridge, MA 02138, USA}
\altaffiltext{2}{Center for Astrophysics Research, University of Hertfordshire, College Lane Campus, Hatfield, 
Hertordshire, AL10 9AB, UK}

\begin{abstract}

We report significant inhomogeneities in the projected two-dimensional (2D) spatial distributions of Low-Mass X-ray 
Binaries (LMXBs) and Globular Clusters (GCs) of the intermediate mass elliptical galaxy NGC4278. In the inner
region of NGC4278, a significant 
arc-like excess of LMXBs extending south of the center at $\sim50^{\prime\prime}$ in the western
side of the galaxy can be associated to a similar over-density of the spatial distribution of red GCs 
from~Brassington et al. (2009). Using a recent catalog of GCs produced by~\cite{usher2013} and covering
the whole field of the NGC4278 galaxy, we have discovered two other significant density structures 
outside the $D_{25}$ isophote to the W and E of the center of 
NGC4278, associated to an over-density and an under-density respectively. We discuss the 
nature of these structures in the context of the similar spatial inhomogeneities discovered in the 
LMXBs and GCs populations of NGC4649 and NGC4261, respectively. These features 
suggest streamers from disrupted and accreted dwarf companions. 

\end{abstract}


\section{Introduction}
\label{sec:intro}

This is the third paper in a series analyzing the two-dimensional (2D) distribution of GC systems in 
elliptical galaxies. We first reported the discovery of spiral-like features in the 2D distribution 
of the GCs of NGC4261 with a K-Nearest-Neighbor (KNN) technique supplemented by Monte Carlo 
simulations~\citep{dabrusco2013a}. We then investigated the GC system of 
NGC4649~\citep{dabrusco2013b}, detecting highly significant spatial inhomogeneities. These results are 
qualitatively in agreement with the discovery of complex GC systems in M87~\citep{strader2011} 
and NGC4365~\citep{blom2012} and reinforce the notion that merging and accretion of satellites 
are important in the evolution of galaxies, as predicted by the $\Lambda$CDM 
theory~\citep[e.g.,][]{dimatteo2005}. 


As is the case for GCs, low-mass X-ray binaries (LMXBs) can also be detected individually 
in elliptical galaxies, with Chandra~\citep[see review,][]{fabbiano2006}, providing another potential 
marker of galaxy evolution. These LMXBs may originate both from the evolution of native stellar binary 
systems, or be formed by dynamical interactions in GCs~\citep{grindlay1984,verbunt1995}. 
In NGC4649, in addition to the inhomogeneities of the GC system, we have also 
seen a non-uniform distribution in the LMXBs~\citep{dabrusco2013b}. Some LMXBs are 
associated with GCs, and therefore, not surprisingly, follow the same spatial distribution, but 
uncorrelated inhomogeneities are also observed in LMXBs in the stellar field of NGC4649. 
These results have potential implications for our understanding of both LMXB formation and the 
general evolution of the parent galaxy. It is therefore important to establish how frequent their 
occurrence may be. 

Here we report the case of a third galaxy, NGC4278, where we have been 
able to detect significant inhomogeneity in both the LMXB and the GC 2D distributions. NGC4278 is 
an elliptical galaxy at a distance of $\sim\!16$ Mpc~\citep{tonry2001}, which was the subject of 
deep {\it Chandra} observations, resulting in a well-characterized catalog of 
LMXBs~\citep[][hereinafter B09]{brassington2009}. This galaxy is an intermediate-size 
elliptical, with stellar mass
smaller than the giant elliptical NGC4261 and NGC4649~\citep[see][for a compilation of 
properties]{boroson2011}, which were the subjects of companion 
studies~\citep{dabrusco2013a,dabrusco2013b}. The spatial 2D 
distribution of LMXBs in NGC4278 (see Figure 1 in~B09) suggests an 
arc-like feature and possible ``streamers'', motivating the present investigation. 

While the entire galaxy was covered with {\it Chandra}, only 
the central region had been observed with HST at the time~\citep{kundu2001}, resulting in
LMXB-GC identifications for the joint {\it Chandra}-HST coverage region. New HST observations 
of a field covering the 
entire NGC4278 and the nearby NGC4283 
have recently become available~\citep{usher2013} (hereinafter U13), complementing the 
GC coverage at larger radii. The positions of the U13 GCs 
suggest the presence of large-scale density structures. In order to get a comprehensive picture of the 
properties of the GCs population in NGC4278, in this paper we investigate 
the projected spatial distribution of both candidate U13 GCs and the GC catalog from~B09.

We discuss the datasets in Section~\ref{sec:data}; we present the results of our analysis of the 
spatial distribution of LMXBs and GCs in the central region of the galaxy NGC4278 and at
larger radii in Section~\ref{sec:results}, and examine their 
astrophysical implications in Section~\ref{sec:discussion}. We draw our conclusions in 
Section~\ref{sec:conclusions}.

\begin{figure}[]   
	\includegraphics[height=8cm,width=8cm,angle=0]{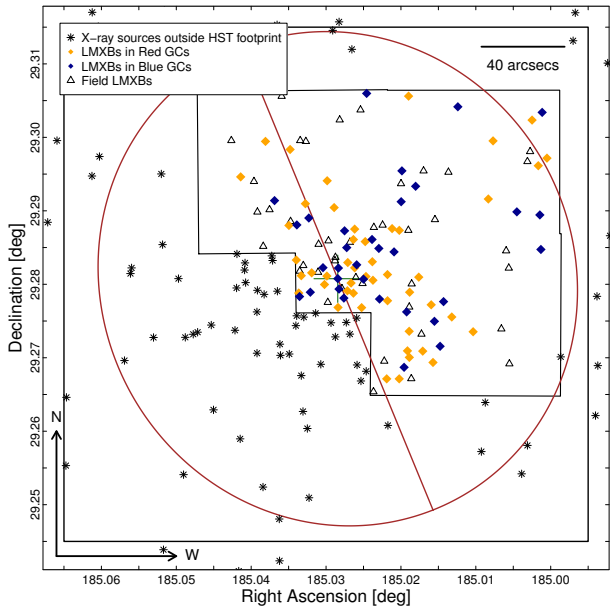}	
	\caption{LMXBs in NGC4278. GC-LMXBs in blue and red GCs 
	are indicated with blue and orange symbols, while field LMXBs are plotted as black triangles.
	The outer solid line represents the {\it Chandra} footprint, and the enclosed area shows the HST 
	footprint used to extract the GC sample. The ellipse represents the $D_{25}$ isophote of NGC4278 
	from~\cite{devaucouleurs1991}. The major axis of NGC4278 is also 
	shown~\citep{devaucouleurs1991}.}
	\label{fig:positions_lmxb}
\end{figure}

\begin{figure*}[]   
	\includegraphics[height=8cm,width=8cm,angle=0]{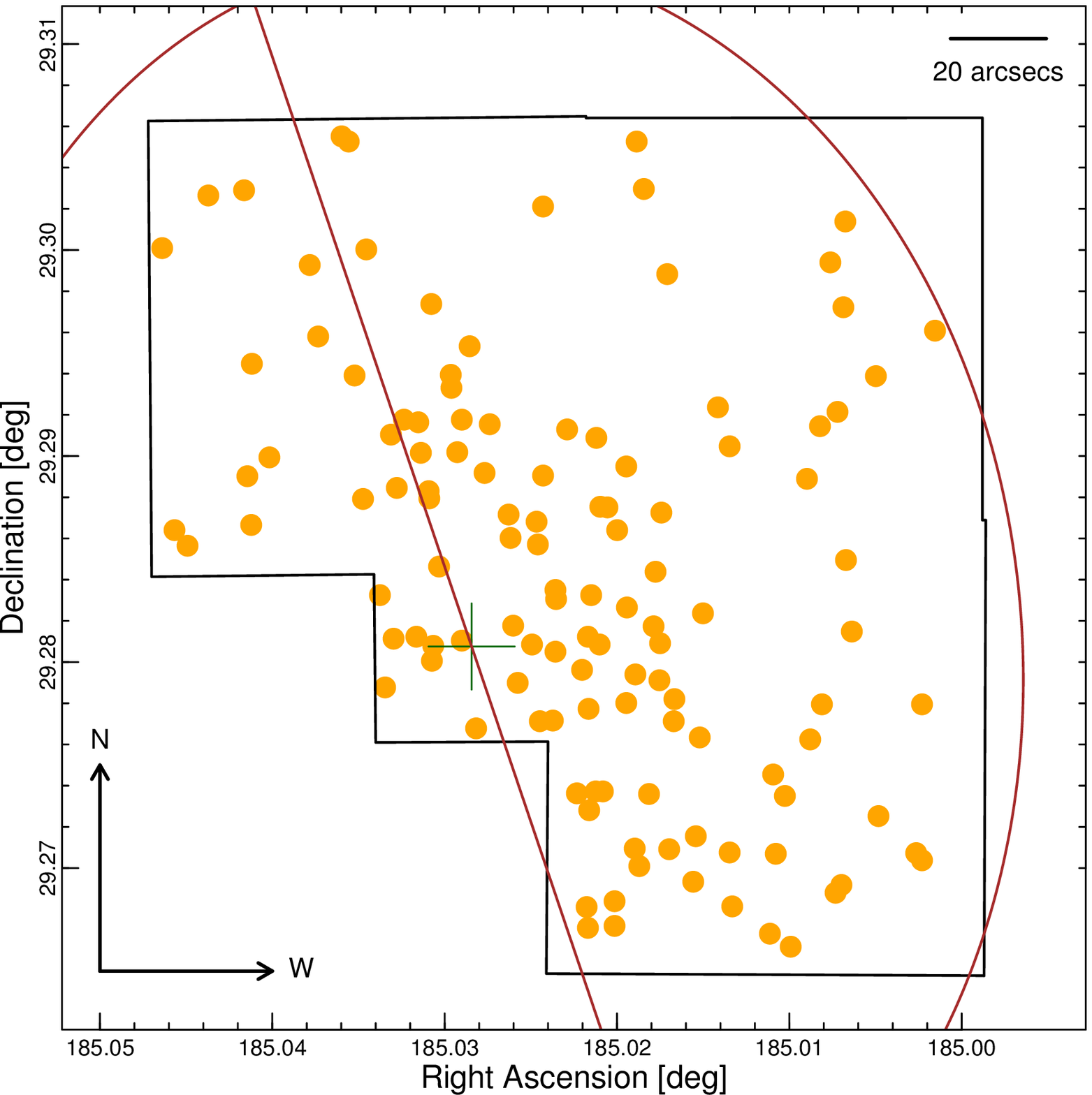}
	\includegraphics[height=8cm,width=8cm,angle=0]{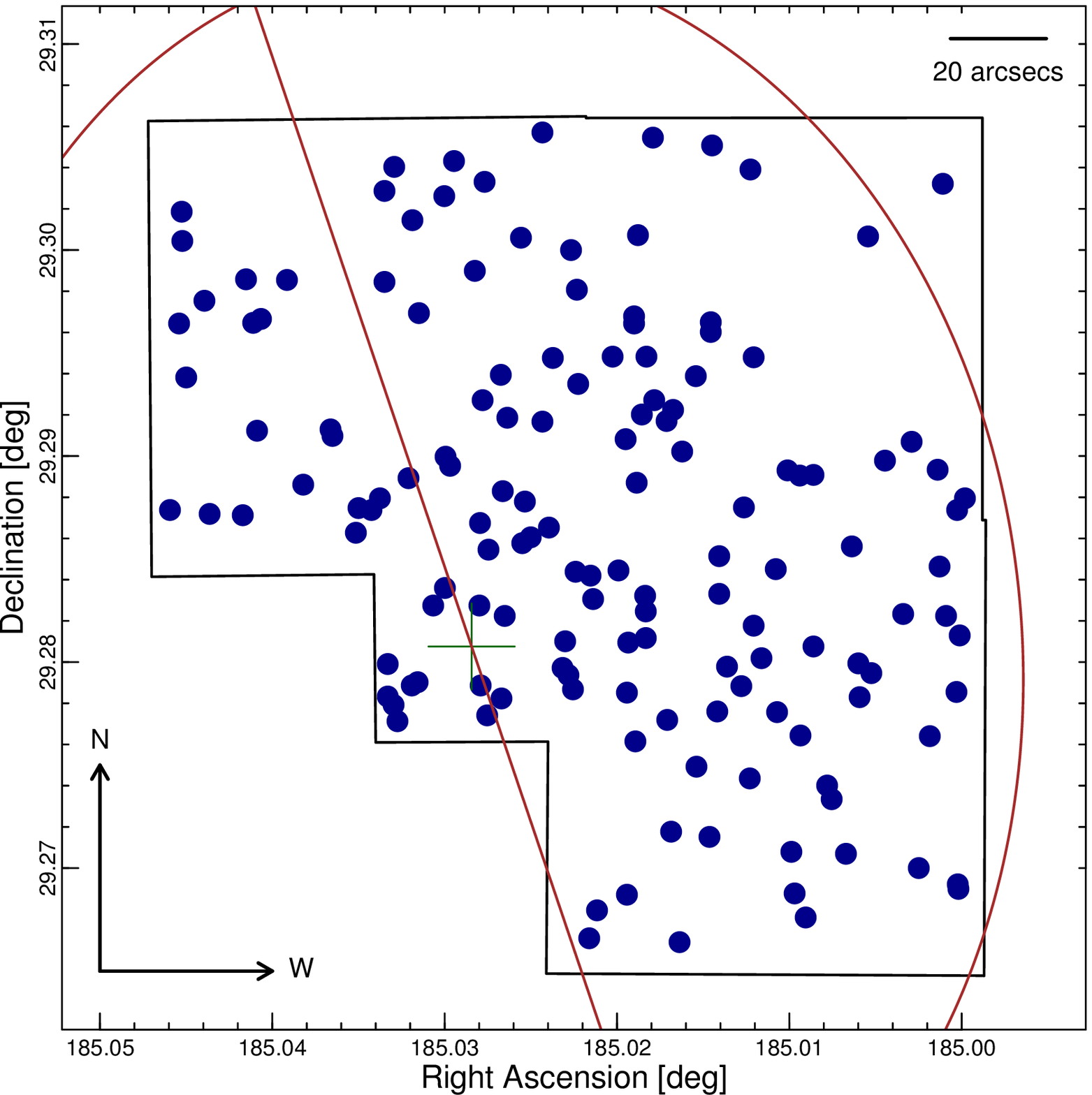}	
	\caption{Positions of red GCs (left) and blue GCs (right) in NGC4278. The ellipse represents 
	the $D_{25}$ isophote of NGC4278 from~\cite{devaucouleurs1991}. 
	The outlines identify the footprint of the HST observations used by~B09 to 
	extract the catalog of GCs. The major axis of NGC4278 the galaxy is also 
	shown~\citep{devaucouleurs1991}.}
	\label{fig:positions_gc}
\end{figure*}

\begin{table}[h]
	\centering
	\caption{Samples of LMXBs and GCs used in this paper.}
	\begin{tabular}{lccc}
	\tableline
			& $N_{\mathrm{tot}}$	&$N_{\mathrm{GCs}}$ &$N_{\mathrm{field}}$ 	\\			
	LMXBs\footnote{In square brackets, the number of LMXBs located inside the $D_{25}$ isophote 
	of NGC4278. In curly brackets, 
	the number of GC-LMXBs associated to blue and red GCs respectively. In parenthesis, 
	the number of LMXBs located outside the HST footprint.}	
			& 	236[171]		        & 80\{34/46\}				    & 	42(114)	\\ 
	\tableline
	\tableline
			& $N_{\mathrm{tot}}$	&$N_{\mathrm{red}}$ &$N_{\mathrm{blue}}$  \\			
	GCs\footnote{Catalog of GCs from~B09, covering the central region of NGC4278.}	
			& 	266			& 	121		  & 	145	 \\
	GCs\footnote{Catalog of GCs from~U13, covering the whole field of the galaxy NGC4278.}	
			& 	716			& 	360		  & 	356	 \\
	\tableline
	\end{tabular}\\
	\label{tab:summary}
\end{table}

\begin{table*}
	\centering
	\caption{Fractions of simulated residual maps of the population of GCs and LMXBs
	with number of extreme pixels (i.e., pixels with density values exceeding the 90-th 
	percentile of the observed pixel residual 
	distribution) larger than the number of observed extreme pixels in the residual map.
	Values in parenthesis refer to the fraction of simulated density maps with at least one 
	group of contiguous extreme pixels as large as the groups of contiguous extreme pixels
	in the observed residual maps.
	These fractions were determined by counting the number of simulated density
	maps with at least one group of contiguous extreme pixels equal or larger than
	the group of contiguous extreme pixels observed over-density regions.}
	\begin{tabular}{lcccc}
	\tableline
	Residuals				&			&			&			\\
		 				&$K\!=\!7$	&$K\!=\!8$	&$K\!=\!9$	\\
	\tableline
	All GCs (red$+$blue)\footnote{GC dataset from~B09}	
						&3.3\%(0.2\%)	&0\%(0\%)	&0\%(0\%)	\\
	Red GCs				&0.2\%(0.3\%)	&0\%(0\%)	&0\%(0\%)	\\
	Blue GCs				&1.2\%(0\%)	&0\%(0\%)	&0\%(0\%)	\\	
	\tableline
	All LMXBs				&12.3\%(5.6\%)		&2.5\%(0\%)		&1.2\%(0\%)		\\
	LMXBs in GCs			&9.8\%(2.1\%)		&1.8\%(0.3\%)		&0.2\%(0\%)		\\
	Field LMXBs			&19.5\%(4.2\%)		&1.4\%(0.1\%)		&0\%(0\%)		\\	
	\tableline
	All GCs (red$+$blue)\footnote{GC dataset from~U13}	
						&1.9\%(0.1\%)	&0\%(0\%)	&0\%(0\%)	\\
	Red GCs				&7.8\%(4.3\%)	&2.5\%(1.2\%)	&0.4\%(0\%)	\\
	Blue GCs				&9.3\%(5.4\%)	&3.9\%(2.0\%)	&0.1\%(0.1\%)	\\	
	
	\tableline	
	\end{tabular}
	\label{tab:statistics}
\end{table*}

\section{Data}
\label{sec:data}

The~B09 catalog lists 236 X-ray sources with 0.3-8.0 keV luminosities ranging 
from $3.5\!\times\!10^{36}$ to $\sim2\!\times\!10^{40}$ ergs s$^{-1}$, and also includes a list of 
GCs from a HST WFPC2 observation~\citep{kundu2001}, partially covering the galaxy 
(Table~\ref{tab:summary}). 
Figure~\ref{fig:positions_lmxb} shows the 2D distribution of these X-ray sources, which suggests 
some spatial features, in particular a region with a 
relative lack of LMXBs in the south between 25$^{\prime\prime}$ and 40$^{\prime\prime}$, followed by an 
arc-like region richer in LMXBs. Thirteen out of a total of 23 LMXBs located in the region of the arc 
are associated with red GCs, with 5 LMXBs associated to blue GCs and 5 field LMXBs. The 2D 
distributions of red (left) and blue (right) GCs separately are shown in Figure~\ref{fig:positions_gc}. 
The red GCs are centrally concentrated and sparse in the N-W corner of the 
observed region. They also seem to be approximately aligned along the major axis of the galaxy, 
with the exclusion of an over-density in the S-W corner 
of the footprint. The blue GCs, on the other hand, are more evenly distributed and do not show
any clear over- or under-density.

\begin{figure*}[]   
	\includegraphics[height=8cm,width=8cm,angle=0]{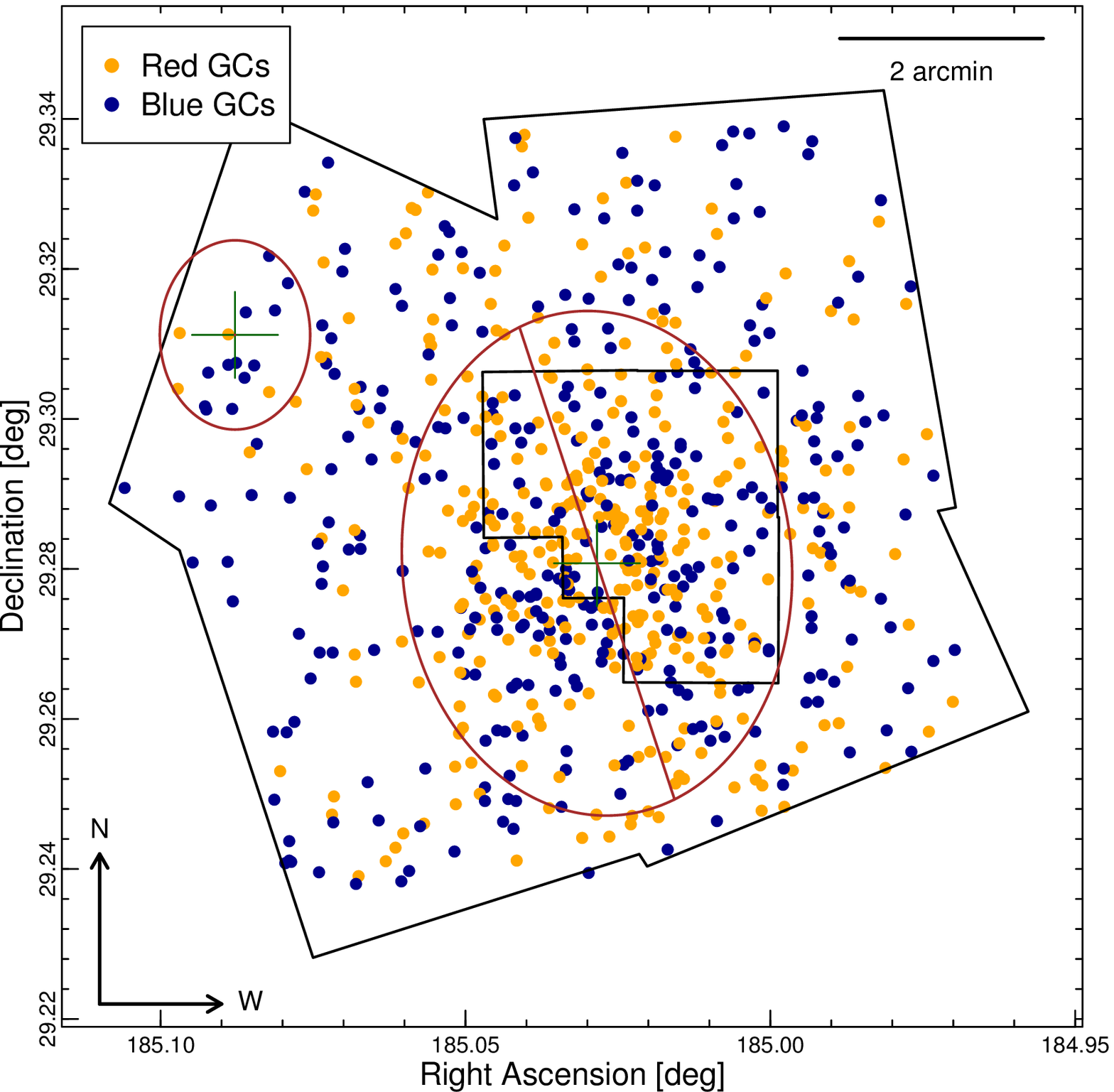}
	\includegraphics[height=8cm,width=8cm,angle=0]{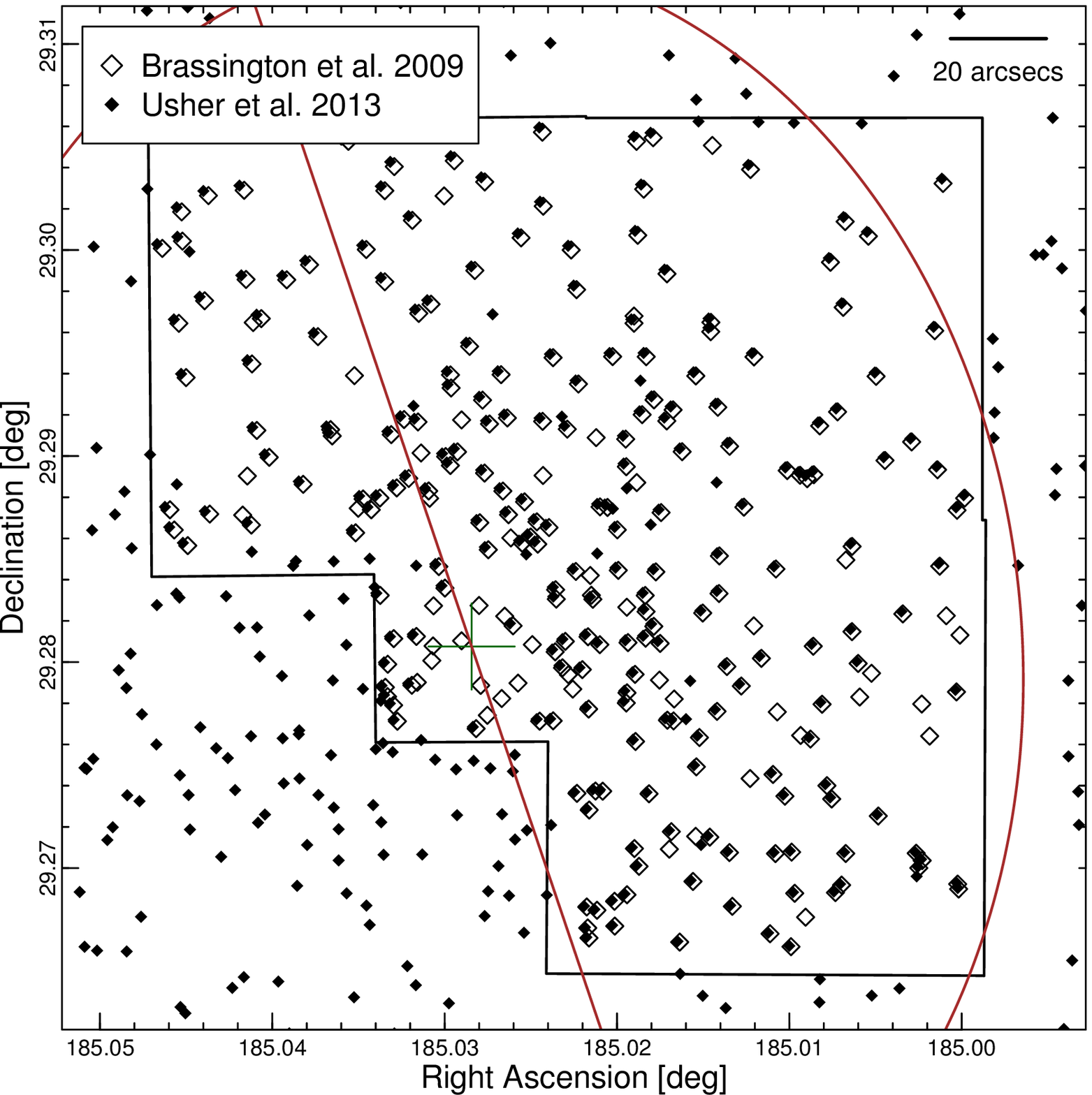}
	\caption{Left: U13 GCs in NGC4278. Blue and red GCs 
	are indicated with blue and orange symbols.The outer solid line represents the
	footprint of the HST ACS observations used to extract the U13 catalog, 
	and the enclosed area shows the HST footprint used to extract the GCs sample 
	from~B09. The ellipses represent the $D_{25}$ isophotes of NGC4278 
	and NGC4283 from~\cite{devaucouleurs1991}. The major axis of NGC4278 is also 
	shown~\citep{devaucouleurs1991}.
	Right: distribution of the GCs extracted from~U13 (solid symbols) and 
	from~B09 (open symbols) in the central region of NGC4278. The black 
	solid line represent the footprint of the HST ACS observations used in~B09.
	The $D_{25}$ elliptical isophote and the major axis of NGC4278 are also shown.}
	\label{fig:positions_gc2}
\end{figure*}

U13 used five archival HST ACS pointings to study the properties of the GC system in 
NGC4278 out to large radial distances. The sample of GC candidate was selected
based on the $g\!-\!z$ color, the $z$ magnitude and the half-light radii of the sources extracted
from the HST images. Contaminants were excluded using a sample of spectroscopically 
confirmed GCs to determine the optimal selection and by visual inspection of all GC candidates 
whose half-light radii exceeded 10 pc (U13). The final catalog of U13 GC candidates, which is
virtually free from contamination, contains 716 sources.~U13  
found that the GC candidates follow a bimodal color distribution, and the red and blue classes can 
be separated using a color threshold $g\!-\!z\!=\!1.078$, leading to 370 red GCs and 346 blue GCs 
(see Table~\ref{tab:summary}). The spatial distribution of U13 GCs
(Figure~\ref{fig:positions_gc2}, left) shows that the red subpopulation
is more centrally concentrated than the blue subpopulation, as commonly observed in elliptical 
galaxies~\cite[see e.g.][]{strader2006}.~U13 also checked the azimuthal distribution 
of GCs in the regions of NGC4278 with complete coverage ($r\!<\!2.5^{\prime}$), finding that the whole 
population of GCs and the two color classes separately are all consistent with a uniformly random 
azimuthal distribution. Figure~\ref{fig:positions_gc2} (right) compares the positions of the GCs 
from~B09 with those of the U13 common area. 
The two distributions are similar on the region of overlap, 
except for the core of the galaxy where the U13 catalog seems to be affected
by a larger incompleteness than the catalog used by~B09. The overall 
agreement between the two GC catalogs shows that hidden systematics are not likely to 
affect our results. As for the LMXB data, any incompleteness effect would have only a radial
dependence, because of the radial worsening of the {\it Chandra} PSF and of the source crowding
in the center. These effects will not affect any 2D inhomogeneity.

\section{Analysis \& Results}
\label{sec:results}

We have analyzed the 2D distributions of LMXBs and GCs, following~\cite{dabrusco2013a}, where 
a detailed explanation of the method used in this paper is given. In brief, for each sample of LMXBs and GCs, 
we have first generated density maps using the K-Nearest Neighbor (KNN) method of~\cite{dressler1980}. 
This density is based on the {\it local} distribution of GCs, i.e. on the distance of the $K$-th closest GCs 
from the points where the density is estimated. We have evaluated the density in each knot of a regular grid 
covering the region of the sky where GCs and LMXBs are found. For a given $K$ value, the point-density is 
estimated as: 

\begin{equation}
	D_{K}\!=\!\frac{K}{A_{D}(d_{K})} 
	\label{eq:knn}
\end{equation}

where $K$ is the index of the nearest neighbor used to calculate the density and $A_{D}\!=\!\pi\cdot d_{K}^2$ 
is the area of the circle with radius equal to the distance of the $K$-th nearest neighbor $d_{K}$. The uncertainty 
on the KNN density scales with the square root of K, so that the relative fractional error is inversely proportional
to the square root of $K$. In other words, the fractional accuracy of the method increases with increasing $K$ 
at the expense of the spatial resolution. In this paper, we have reconstructed the density maps of the 
samples considered using values of $K$ ranging from 2 to 10. While the ``optimal'' value of $K$ depends
on the typical size (expressed as number of members) of the spatial structures that we want to highlight, this
number cannot be determined {\it a priori}. For this reason, we consider an interval of ``reasonable'' $K$ values 
that are roughly proportional to the average density of sources of the samples for which the density maps are
reconstructed. Moreover, large values of $K$ would result in over- and under-density structures that are 
smoothed over large areas and that have lost part of the spatial information they carry. 

We produced residual maps, by subtracting from the observed density maps a 
smooth distribution based on the observed radial number density profiles. Using the smooth distribution 
as seed for Monte Carlo simulations, we have then estimated the statistical significance of any inhomogeneities 
apparent in the residual map. Given the relatively sparse data we have used 5000 simulations in each case. 
Table~\ref{tab:statistics} shows the percentage
of simulated distributions of LMXBs and GCs with ``extreme'' number of pixels, i.e. with density values 
exceeding those of the 90\% of pixels in the observed density map.
The fraction of spatially clustered simulated distributions of GCs and LMXBs from~B09 becomes 
negligible for $K\!=\!8$ and $K\!=\!9$ respectively, while the same fractions for all GCs and red/blue U13 GCs
become consistent with zero for $K\!=\!7$ and $K\!=\!9$ respectively.

\begin{figure*}[]
	\includegraphics[height=8cm,width=8cm,angle=0]{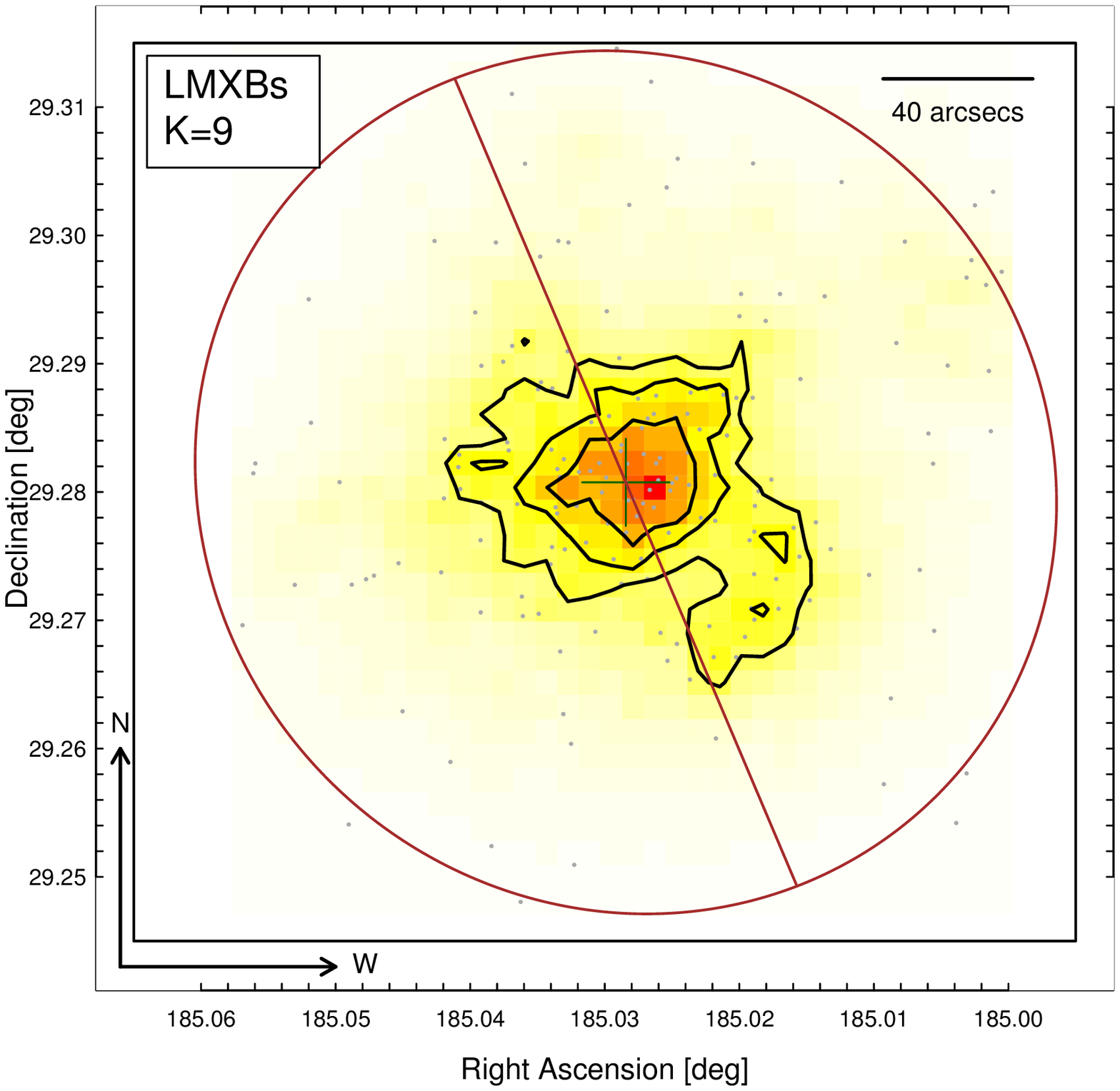}
	\includegraphics[height=8cm,width=8cm,angle=0]{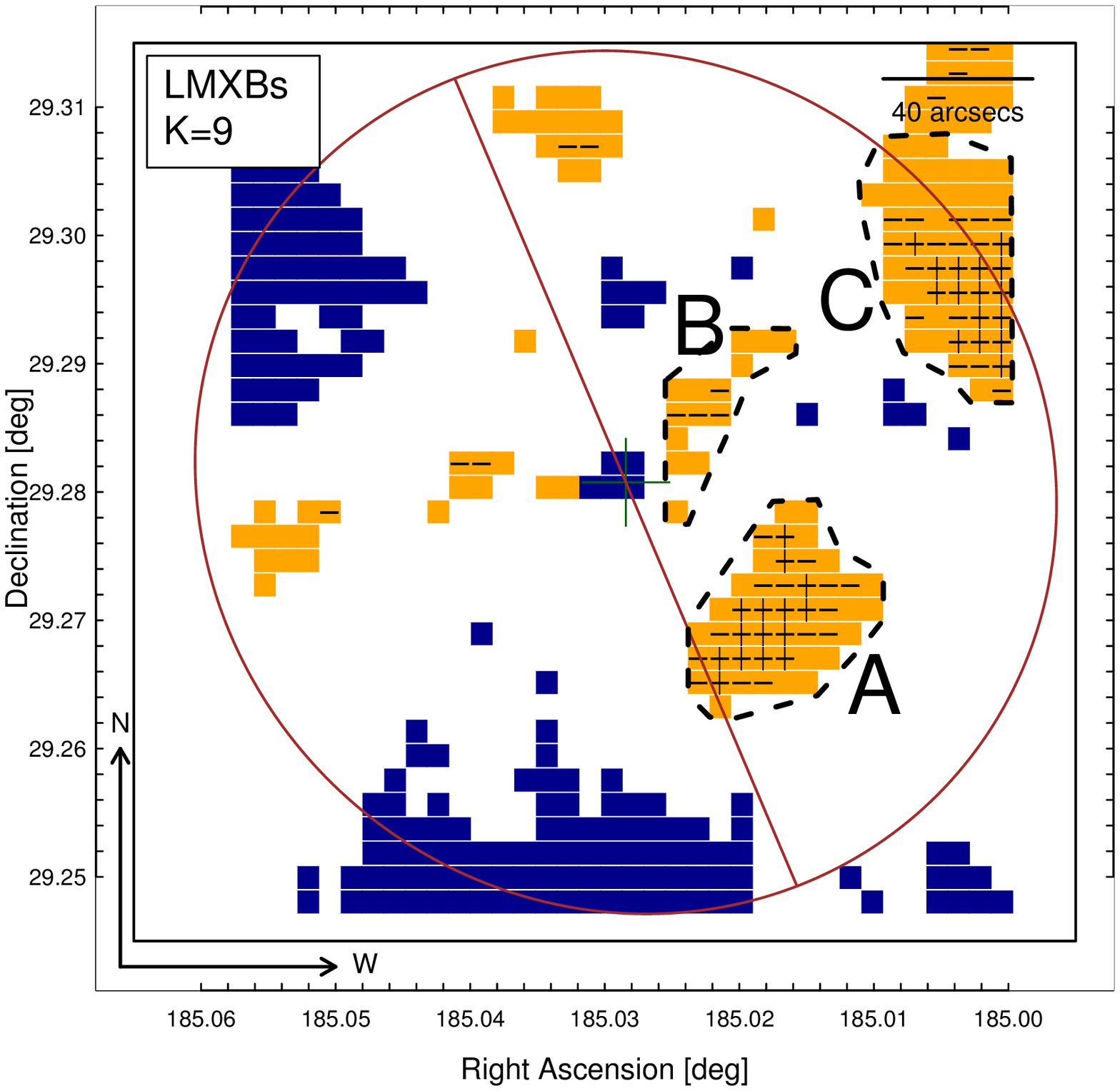}
	\caption{Left: $K\!=\!9$ density maps of the sample of LMXBs in NGC4278. Arbitrary isodensity contours 
	show the higher-density regions in each map. Right: Positions of the $K\!=\!9$ residuals with significance 
	larger than 
	1$\sigma$, 2$\sigma$ and 3$\sigma$ obtained from the residual maps derived from the 
	distribution of the whole catalog of LMXBs in NGC4278. All the negative residuals (blue pixels)
	have significance between 1 and 2 $\sigma$. Positive (orange) pixels $\!>\!2\sigma$ are indicated 
	with a horizontal line; $\!>\!3\sigma$ with a cross.} 
	\label{fig:2dmapsNGC4278_lmxb}
\end{figure*}

\begin{figure*}[]
	\includegraphics[height=8cm,width=8cm,angle=0]{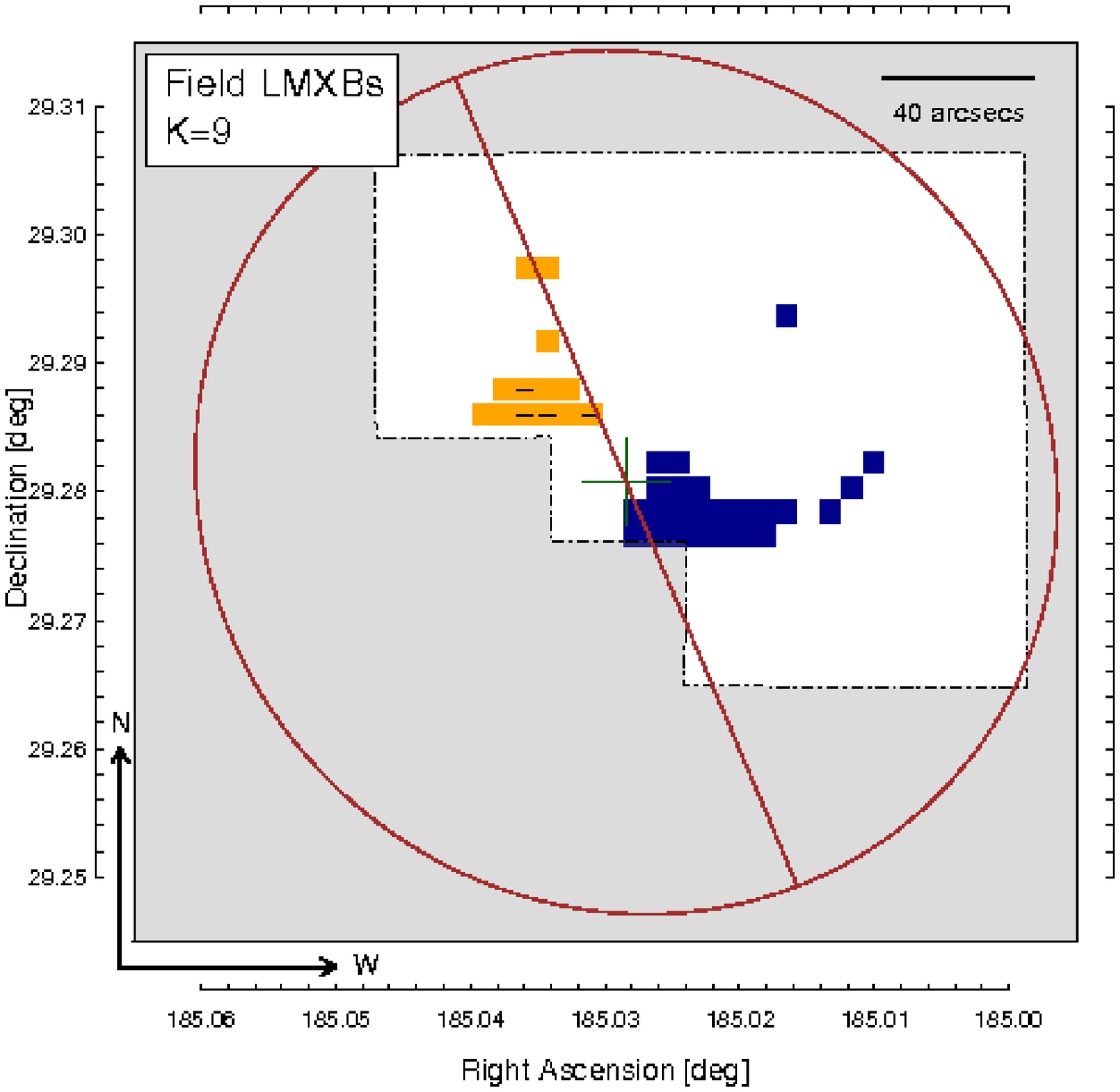}
	\includegraphics[height=8cm,width=8cm,angle=0]{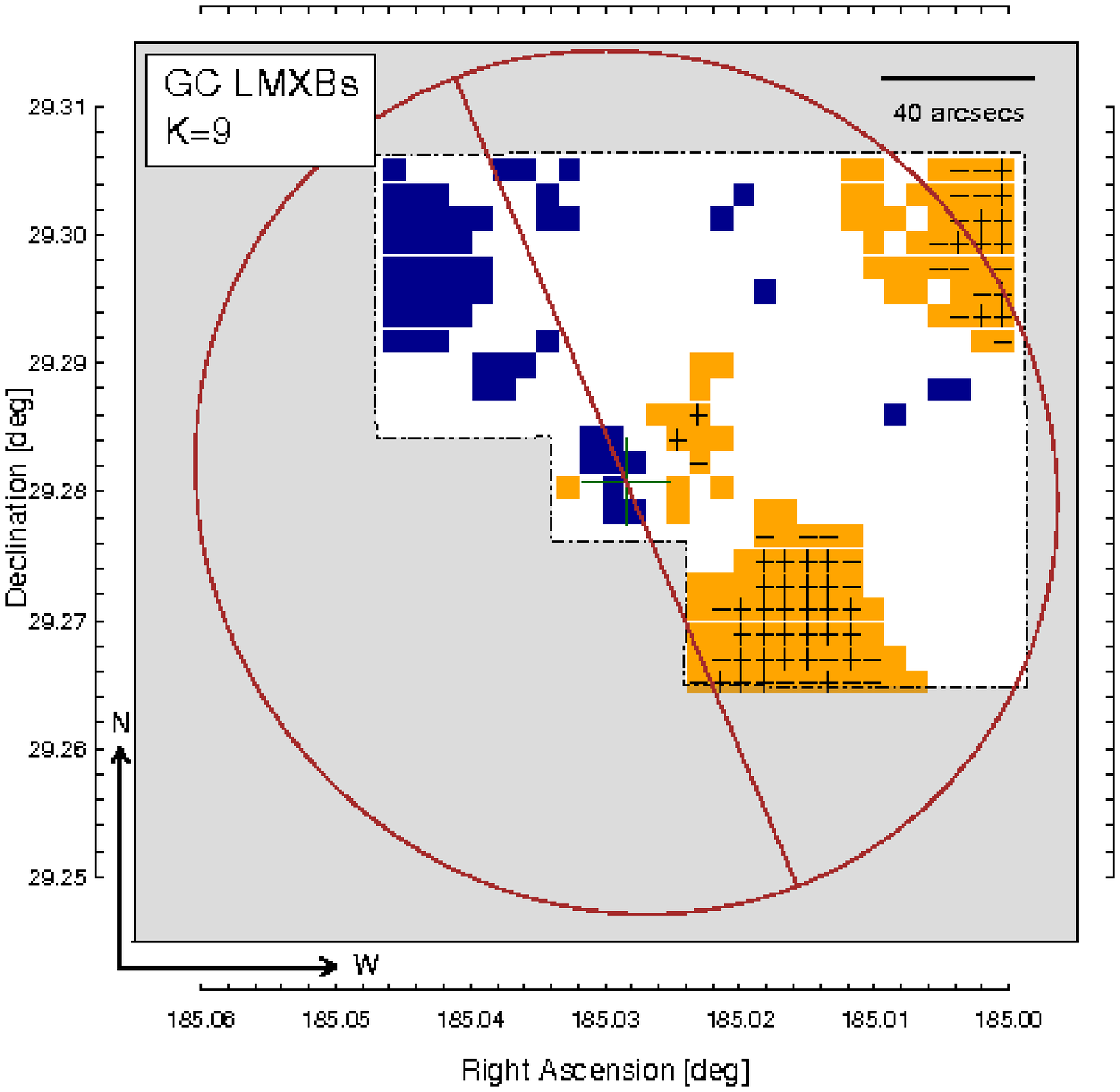}\\
	\caption{Positions of the $\!>\!1\sigma$ residuals for field (left) and GCs-LMXBs
	(right) samples of LMXBs for $K\!=\!9$. Positive (orange) pixels $\!>\!2\sigma$ are indicated 
	with a horizontal line; $\!>\!3\sigma$ with a cross. 
	In all plots the $D_{25}$ elliptical isophote 
	of NGC4278 and the footprints of the {\it Chandra} (and HST, for the field-GC classes) 
	observations used to extract the GCs and LMXBs catalogs respectively are shown for reference.}
	\label{fig:sigma_pixels_maps_lmxb}
\end{figure*}

\begin{table}[h]
	\centering	
	\caption{Significance of the main over-density structures of the LMXBs (``A'', ``B'', ``C'') 
	(Figure~\ref{fig:2dmapsNGC4278_lmxb}, right panel), GCs from~(B09) 
	(``D'', ``E'', ``F'') (Figure~\ref{fig:2dmapsNGC4278_gc}, mid and right panels) and U13 GCs 
	(``G'', ``H'', ``I'', ``L'', ``M'') (Figure~\ref{fig:2dmapsNGC4278_gc_new}, 
	lower panels) residual maps. In parenthesis, the significance of the groups of pixel in the~(B09) 
	GCs residual maps corresponding to the over-densities defined in the~(B09) LMXBs 
	significance map.}
	\begin{tabular}{lcccc}
	\tableline
			& 	LMXBs 		&	All GCs			& 	Red GCs 			& 	Blue GCs  		\\	
	\tableline				
		A	& $\sim\!8\sigma$	& ($\sim\!0\sigma$)		& ($\sim\!6\sigma$) 		& ($\sim\!0\sigma$)		\\
		B	& $\sim\!5\sigma$ 	&($\sim\!1.5\sigma$)		& ($\sim\!0\sigma$) 		& ($\sim\!2\sigma$) 		\\
		C	& $>\!10\sigma$	& ($\sim\!0\sigma$)		& ($\sim\!0\sigma$) 		& ($\sim\!0\sigma$)		\\
		D	&	-			&		-			& $\sim\!7.5\sigma$		&	-				\\
		E	&	-			&		-			& $\sim\!2.5\sigma$		&	-				\\
		F	&	-			&		-			&		-			& $\sim\!4\sigma$		\\		
		G	&	-			& $\sim\!10\sigma$		& $>\!10\sigma$		& $\sim\!10\sigma$		\\		
		H	&	-			& $>\!6\sigma$			& $\sim10\sigma$		& $>\!5\sigma$			\\		
		I	&	-			& $\sim\!5\sigma$		& $>\!10\sigma$	 	& $\sim\!4\sigma$		\\
		L	&	-			& $<\!1.5\sigma$		& $\sim\!2.7\sigma$		& $\sim\!0\sigma$		\\
		M	&	-			& $\sim\!4\sigma$		& $<\!1.5\sigma$		& $\sim\!3.2\sigma$		\\		
	\tableline					
	\end{tabular}
	\label{tab:significance}
\end{table}

\subsection{LMXBs and GCs in the central region}
\label{subsec:central}

Here we discuss the results of the residual maps obtained using the~B09 LMXBs and GC samples. 
Figure~\ref{fig:2dmapsNGC4278_lmxb} shows the results for the LMXB distribution. The density map (left) 
shows that LMXBs are centrally concentrated, with one significant over-density located along the southern 
major axis and extending west-ward. The existence of this over-density is confirmed
by the pixel map of residuals with significance larger than $1\sigma$ (right, 
Figure~\ref{fig:2dmapsNGC4278_lmxb}). In particular, the ``arc'' at $R\!\sim\!1^{\prime}$ 
in the SW quadrant, indicated as ``A'' in Figure~\ref{fig:2dmapsNGC4278_lmxb}, is $\sim\!8\sigma$
significant. We calculate the significance of a each single feature by counting the fraction of 
simulated distributions of LMXBs or GCs with groups of spatially clustered pixels with similar average
significance. We found that the radial ``streamer'' extending to the 
NW of the center of the galaxy, indicated as ``B'' in the 
Figure~\ref{fig:2dmapsNGC4278_lmxb}, has $5\sigma$ significance. The other two 
apparent ``streamers'' point towards E and N, but their statistical significance is low.  	
Repeating the analysis for GC and field LMXBs (within the HST footprint) separately 
(Figure~\ref{fig:sigma_pixels_maps_lmxb}), we only find an excess corresponding to the ``arc''
with the red GCs. 

\begin{figure*}[]
	\includegraphics[height=6cm,width=6cm,angle=0]{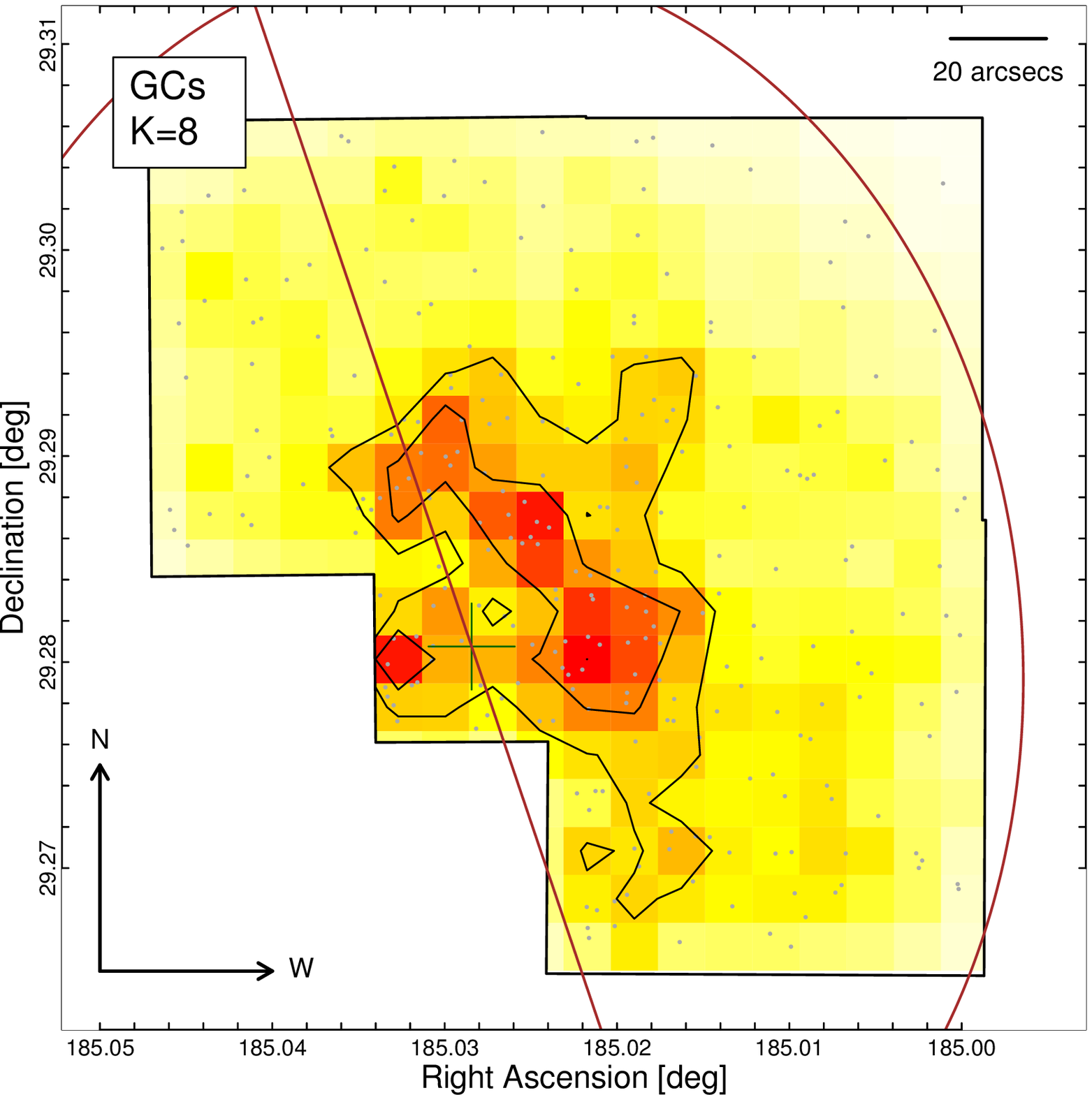}	
	\includegraphics[height=6cm,width=6cm,angle=0]{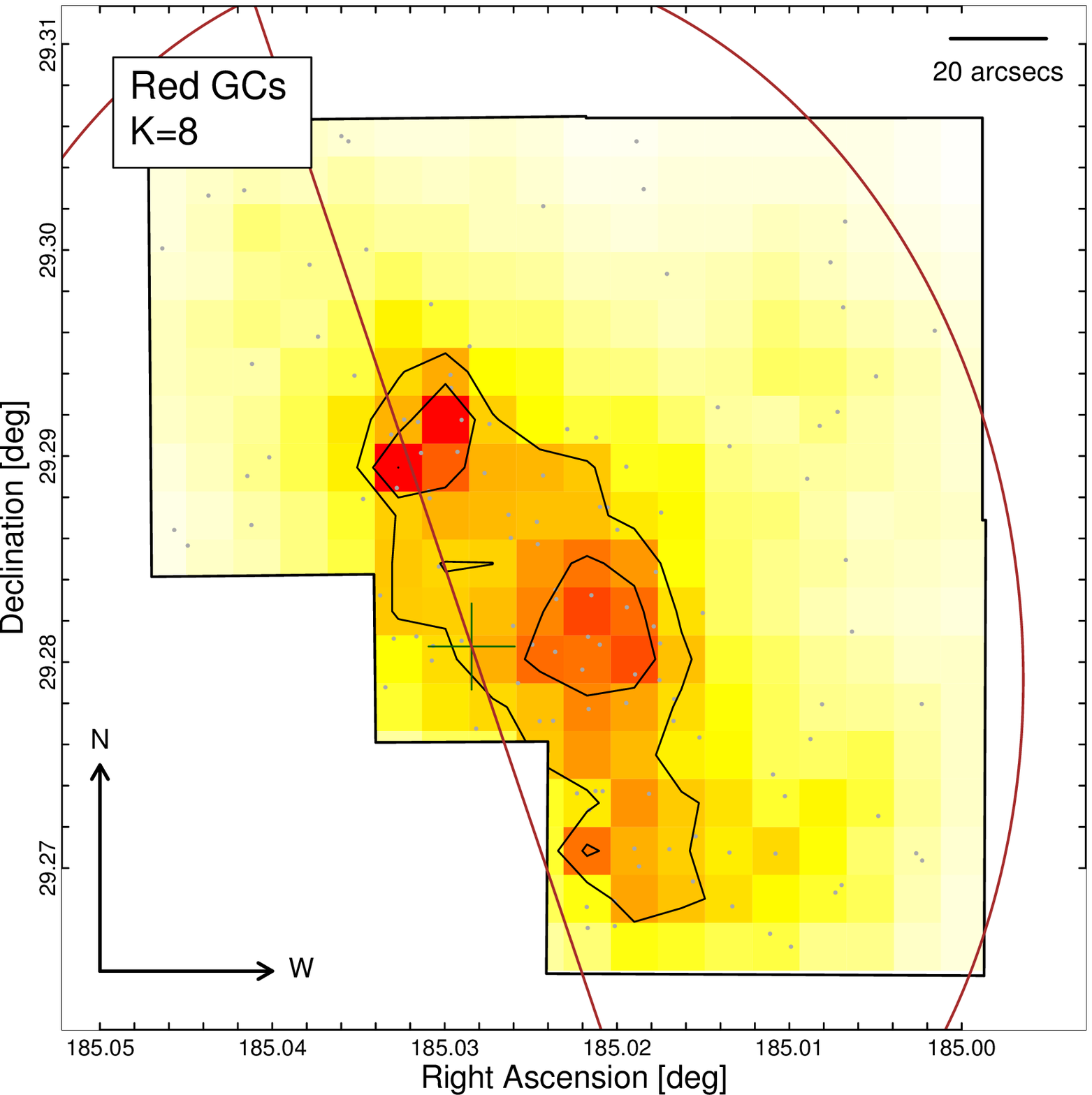}
	\includegraphics[height=6cm,width=6cm,angle=0]{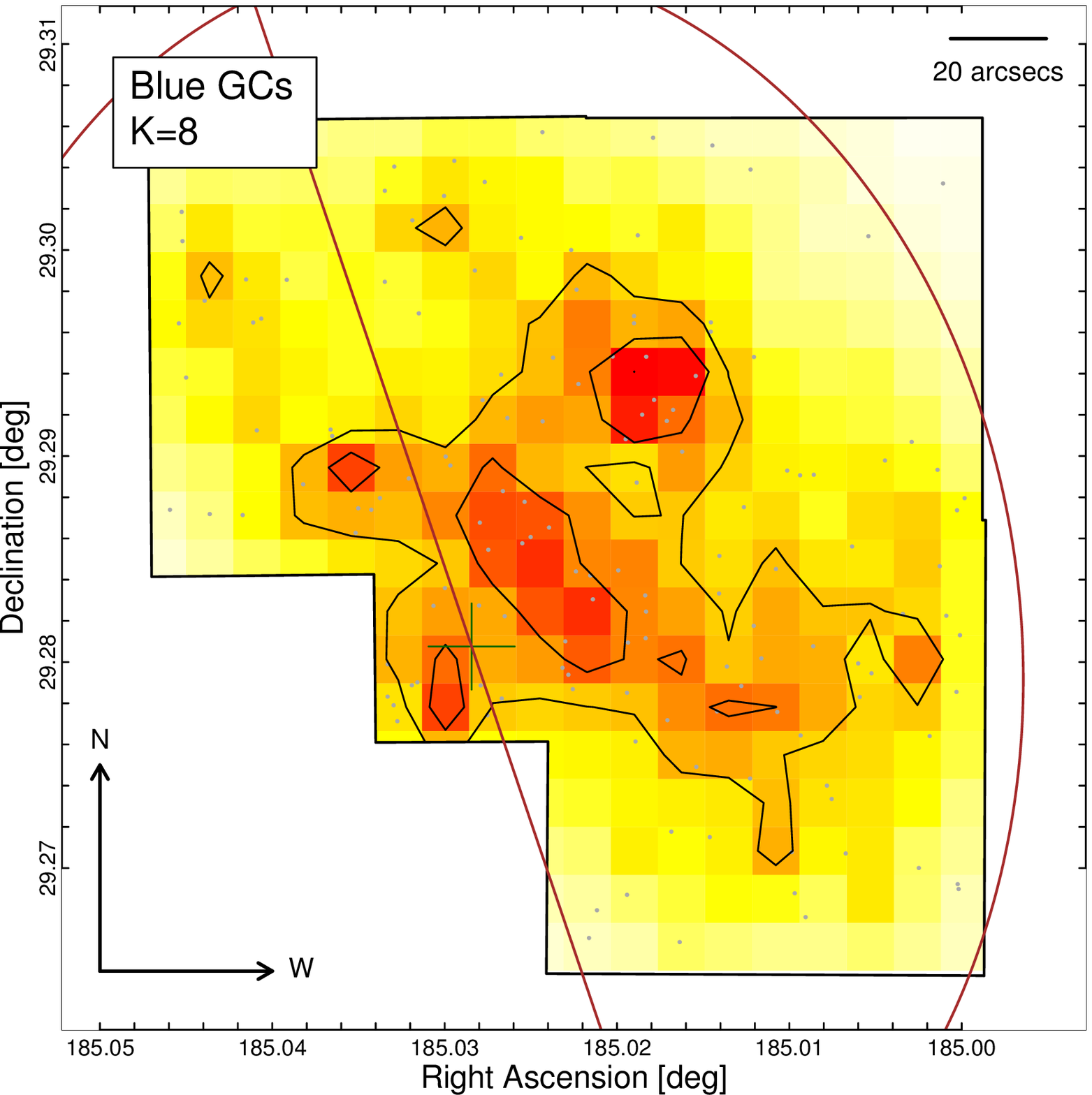}\\		
	\includegraphics[height=6cm,width=6cm,angle=0]{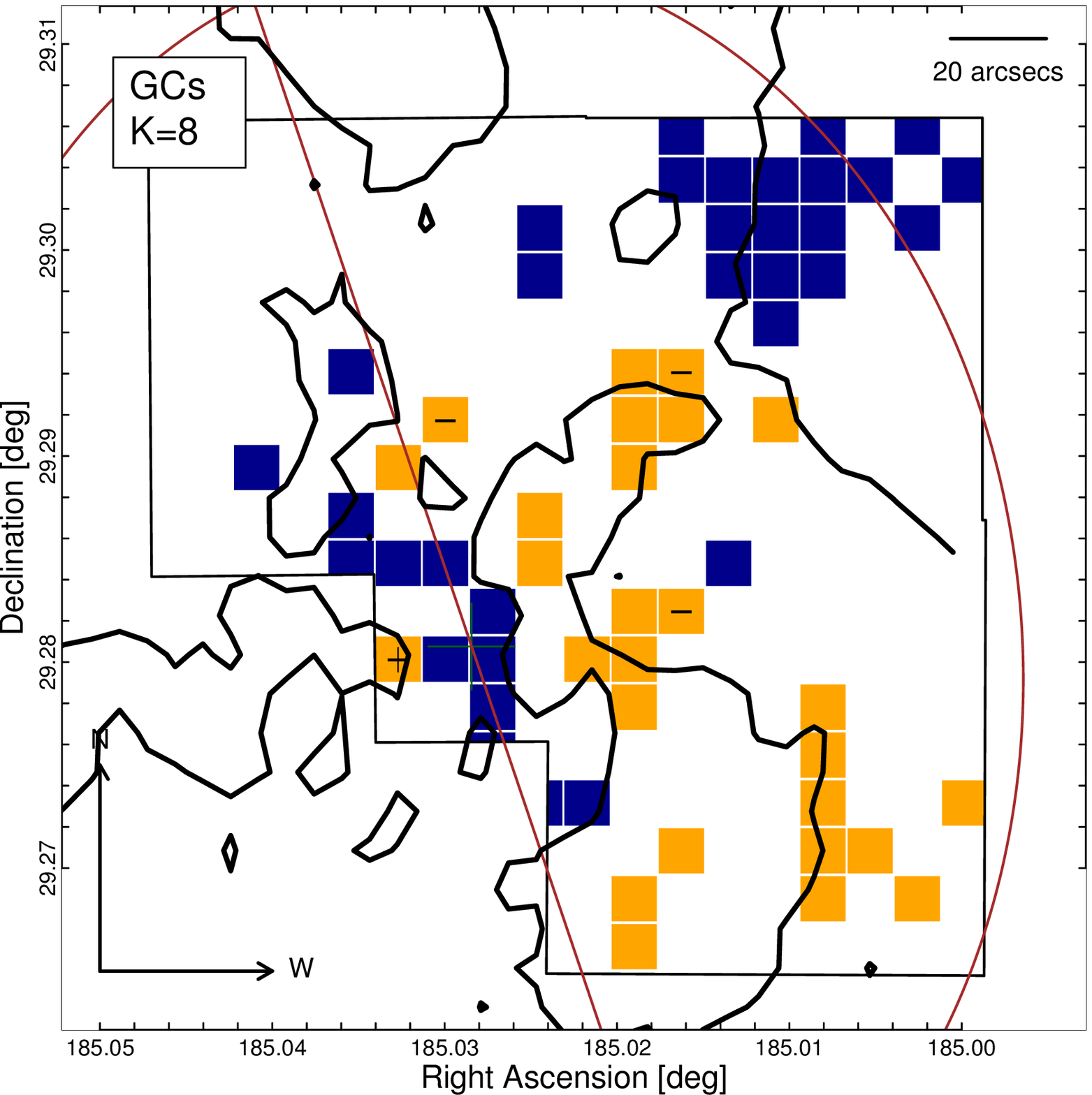}
	\includegraphics[height=6cm,width=6cm,angle=0]{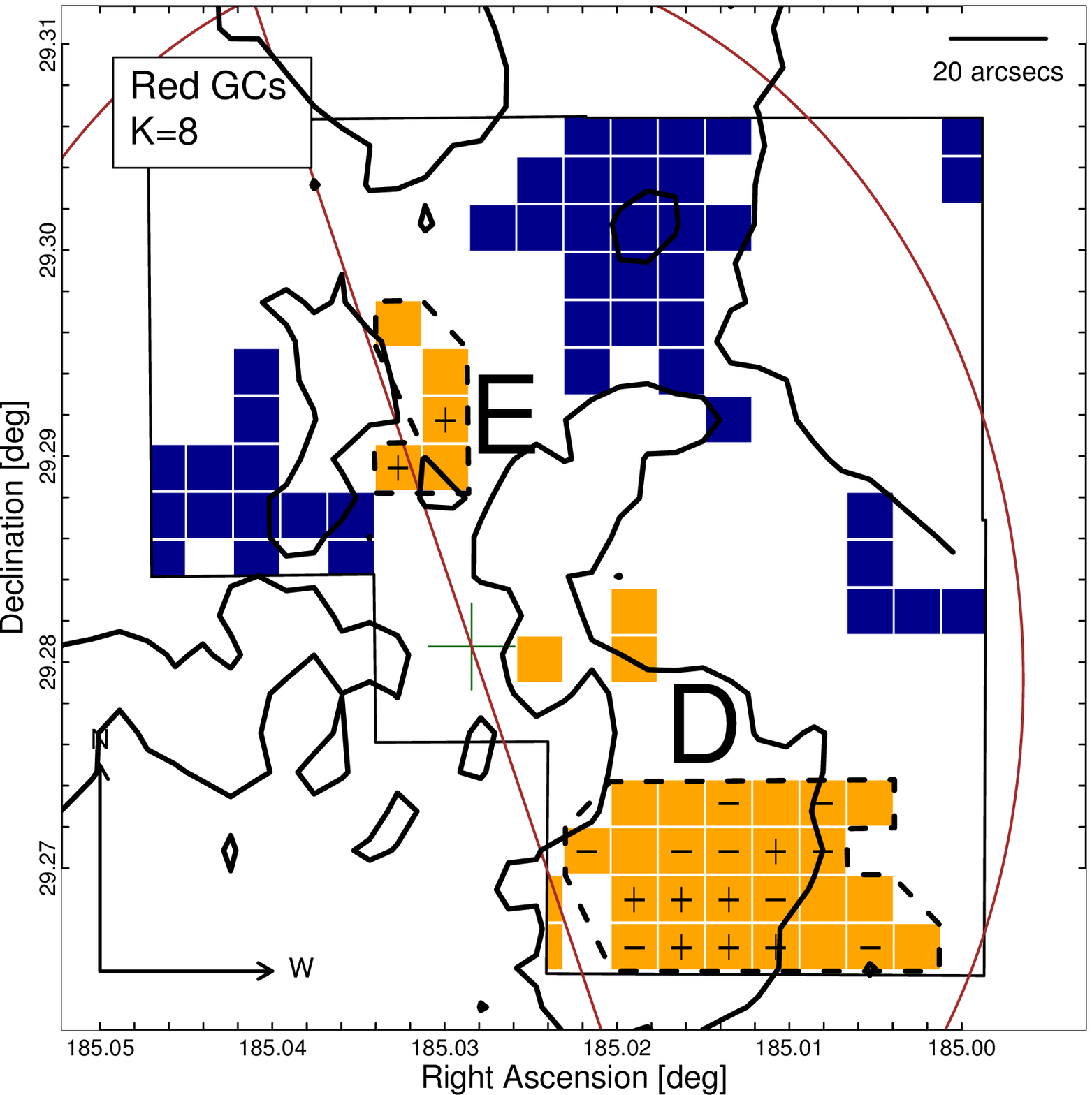}	
	\includegraphics[height=6cm,width=6cm,angle=0]{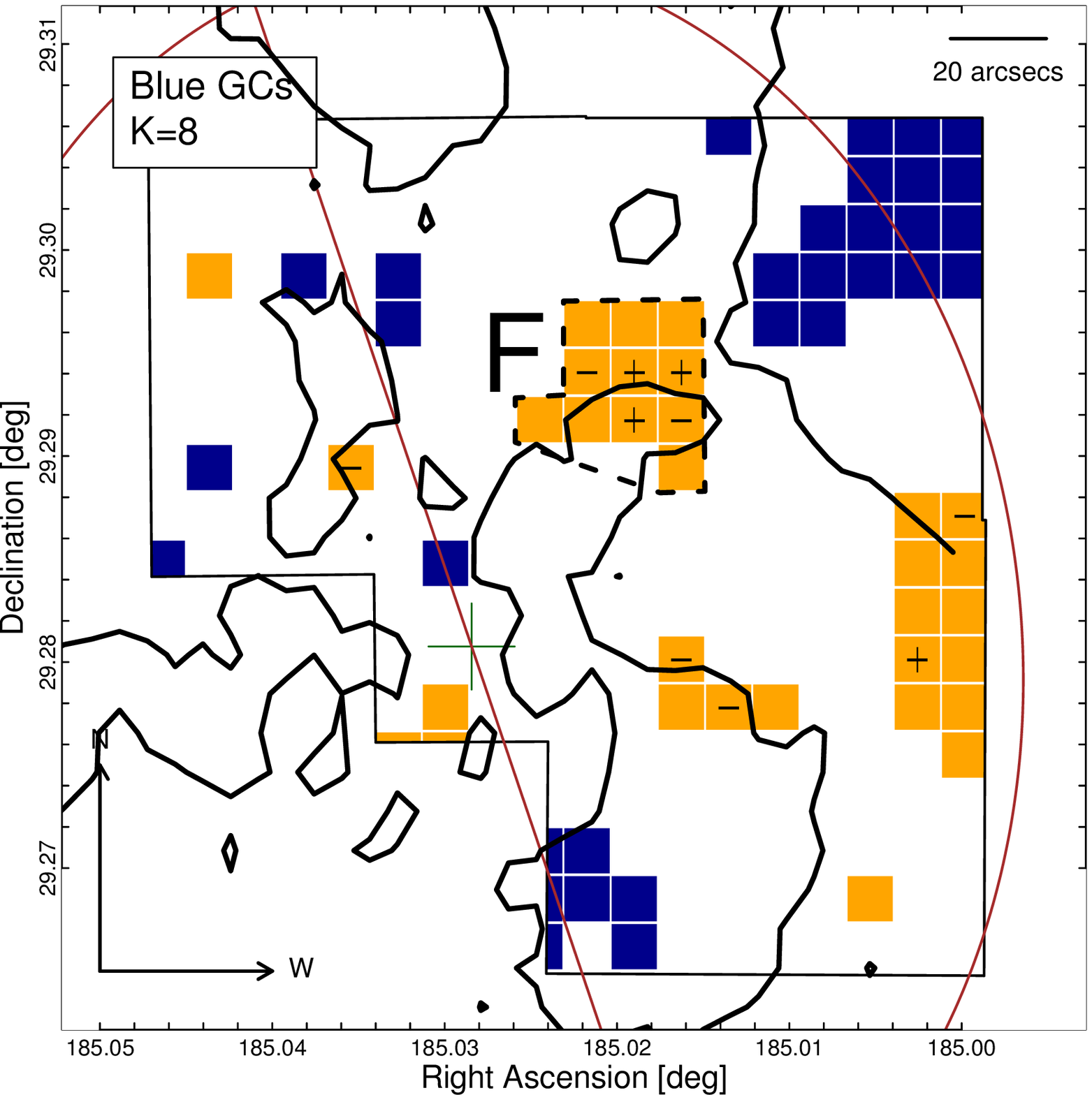}\\	
	\caption{Upper panels: from left to right, $K\!=\!8$ density maps of all, red and blue GCs of the NGC4278. 
	Arbitrary isodensity contours show the higher-density regions. 
	Lower panels: from left to right, $K\!=\!8$ residual maps of all, red and blue GCs of the NGC4278 sample. 
	Pixels are color-coded according to the number of $\sigma$ the pixel deviates from the average. Darker colors indicate larger
	residuals: blue, negative; orange, positive. The small $+, -$ and $=$ signs within each 
	pixel indicate positive, negative or null residuals respectively. The footprint of the HST observations
	used to extract the catalog of GCs and the $D_{25}$ ellipses of NGC4278 and the iso-density contours from the 
	residual maps of the LMXBs distribution are also shown.}
	\label{fig:2dmapsNGC4278_gc}
\end{figure*}

Distinct 2D features can also be seen in the GC distribution, at higher overall statistical 
significance. Figure~\ref{fig:2dmapsNGC4278_gc} 
shows the density maps (upper panels) and residual maps (lower panels) for all, red and 
blue GCs respectively. The residual map of the entire (red $+$ blue) GCs sample does not
displays significant inhomogeneities, while the residual maps for red and blue GCs (obtained for
for $K\!=\!8$) show clear localized 
over-densities. As already discussed in Section~\ref{sec:data}, the red GCs follow the 
direction of the major axis 
of NGC4278, while blue GCs tend to cluster at larger radial distances. Significant 
excesses of red GCs can be seen in the mid-lower panel in 
Figure~\ref{fig:2dmapsNGC4278_gc} in the area of the LMXB arc ($>\!7.5\sigma$, ``D''), 
and, with a lower significance, just north of the region of the NW LMXB streamer ($>\!2.5\sigma$, ``E''). 
The significance map for blue GCs (lower-right panel in Figure~\ref{fig:2dmapsNGC4278_gc})
shows one moderately significant over-density ($\sim\!4\sigma$, ``F'') overlapping the position of 
LMXBs ``streamer'' indicated as ``B'' in Figure~\ref{fig:2dmapsNGC4278_lmxb}. 
A summary of the significances of the main over-density structures seen in the LMXBS and GCs 
residual maps (right plot in Figure~\ref{fig:2dmapsNGC4278_lmxb} and lower mid and right plots in 
Figure~\ref{fig:2dmapsNGC4278_gc} respectively) is presented in Table~\ref{tab:significance}. 


\subsection{The GC distribution at larger radii}
\label{subsec:field}

\begin{figure*}[]
	\includegraphics[height=6cm,width=6cm,angle=0]{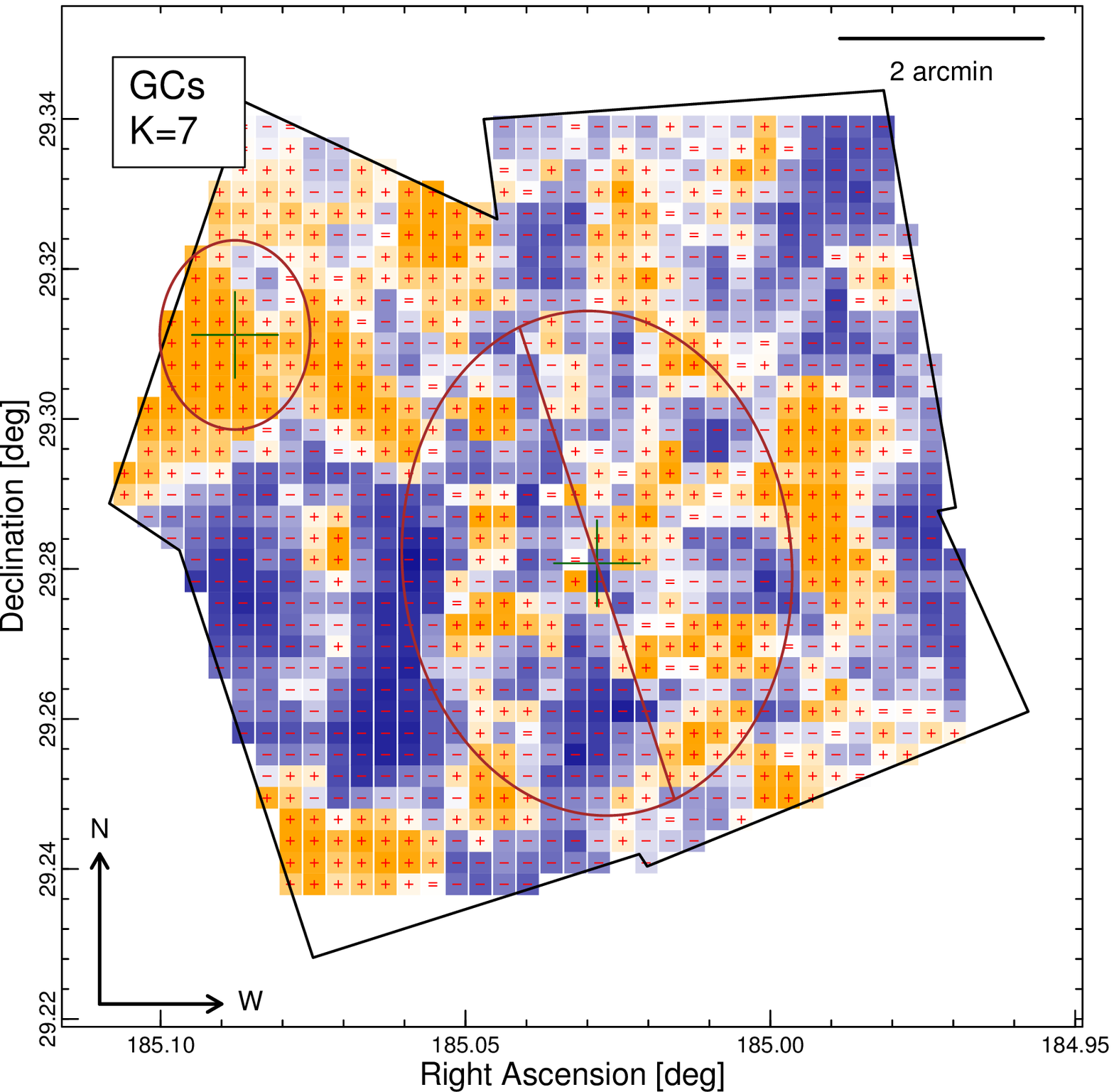}	
	\includegraphics[height=6cm,width=6cm,angle=0]{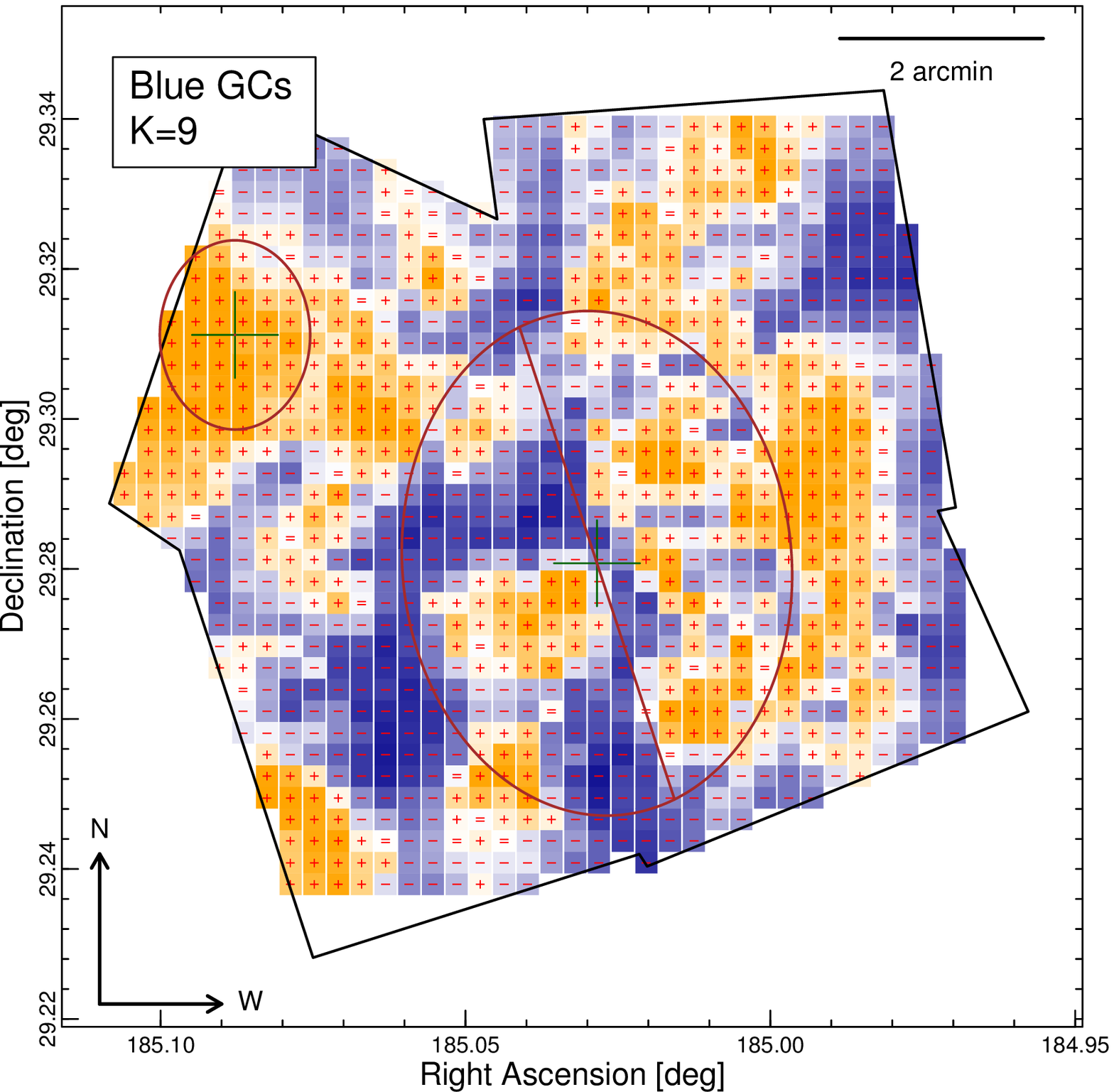}
	\includegraphics[height=6cm,width=6cm,angle=0]{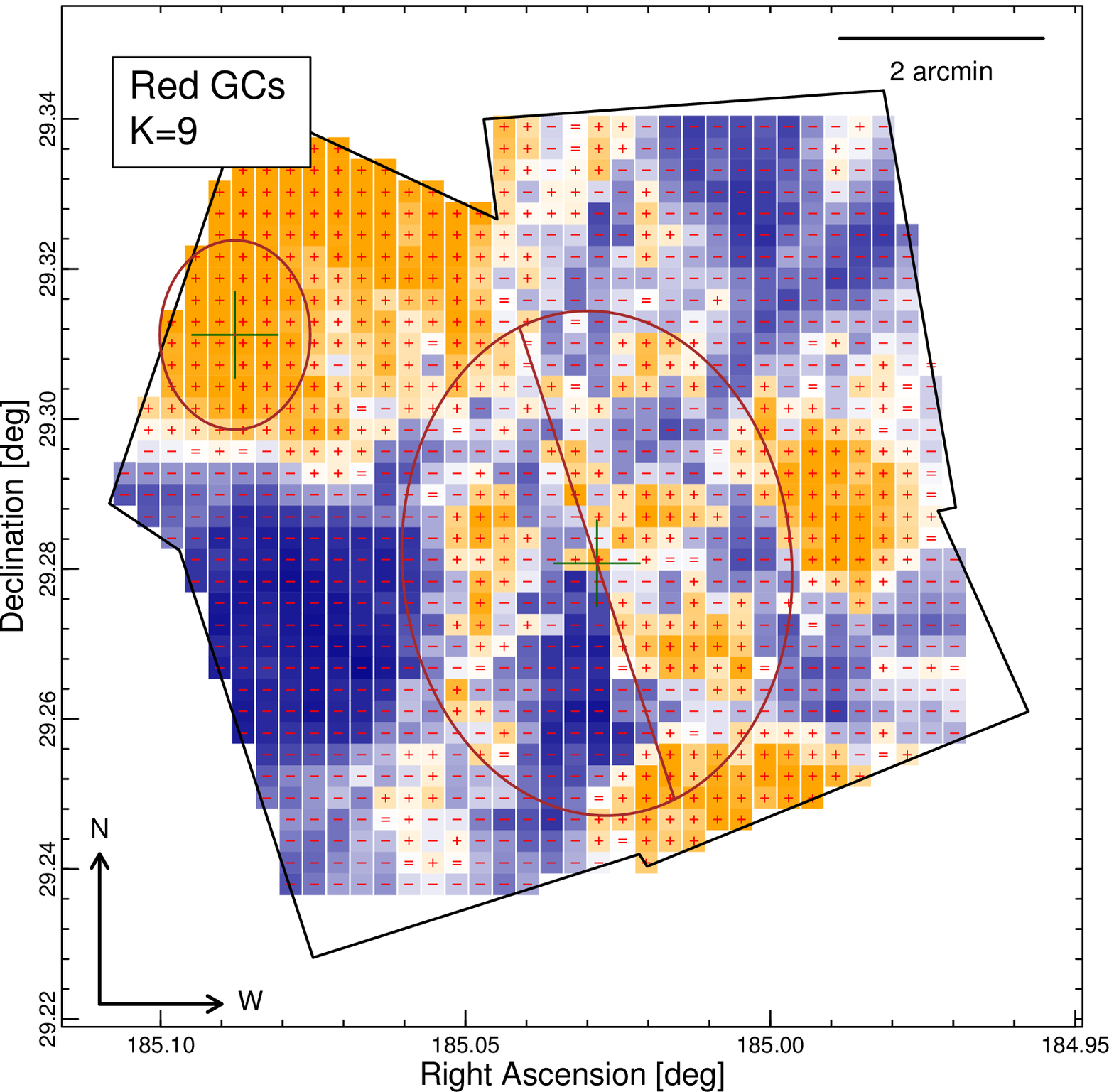}\\		
	\includegraphics[height=6cm,width=6cm,angle=0]{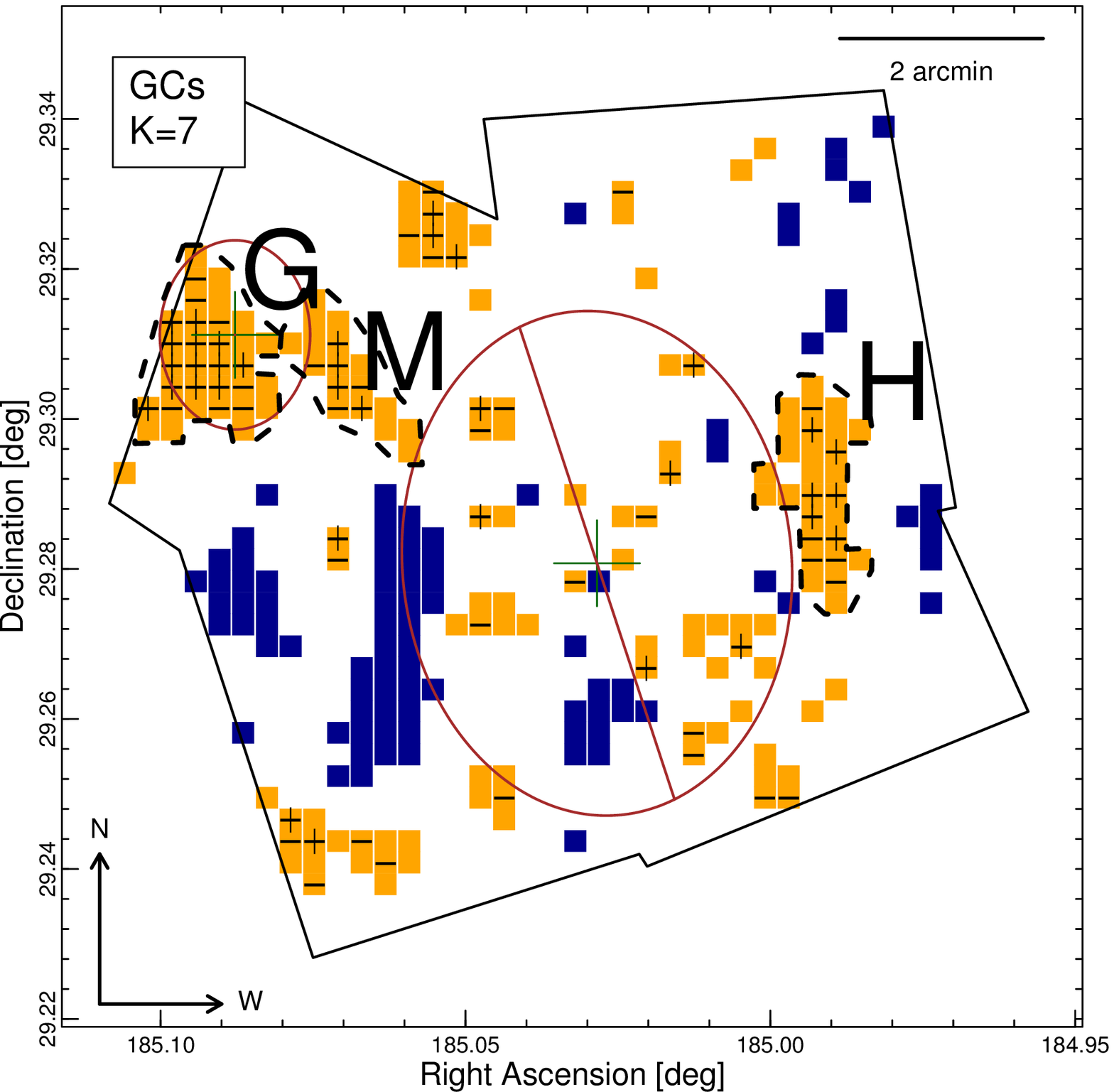}
	\includegraphics[height=6cm,width=6cm,angle=0]{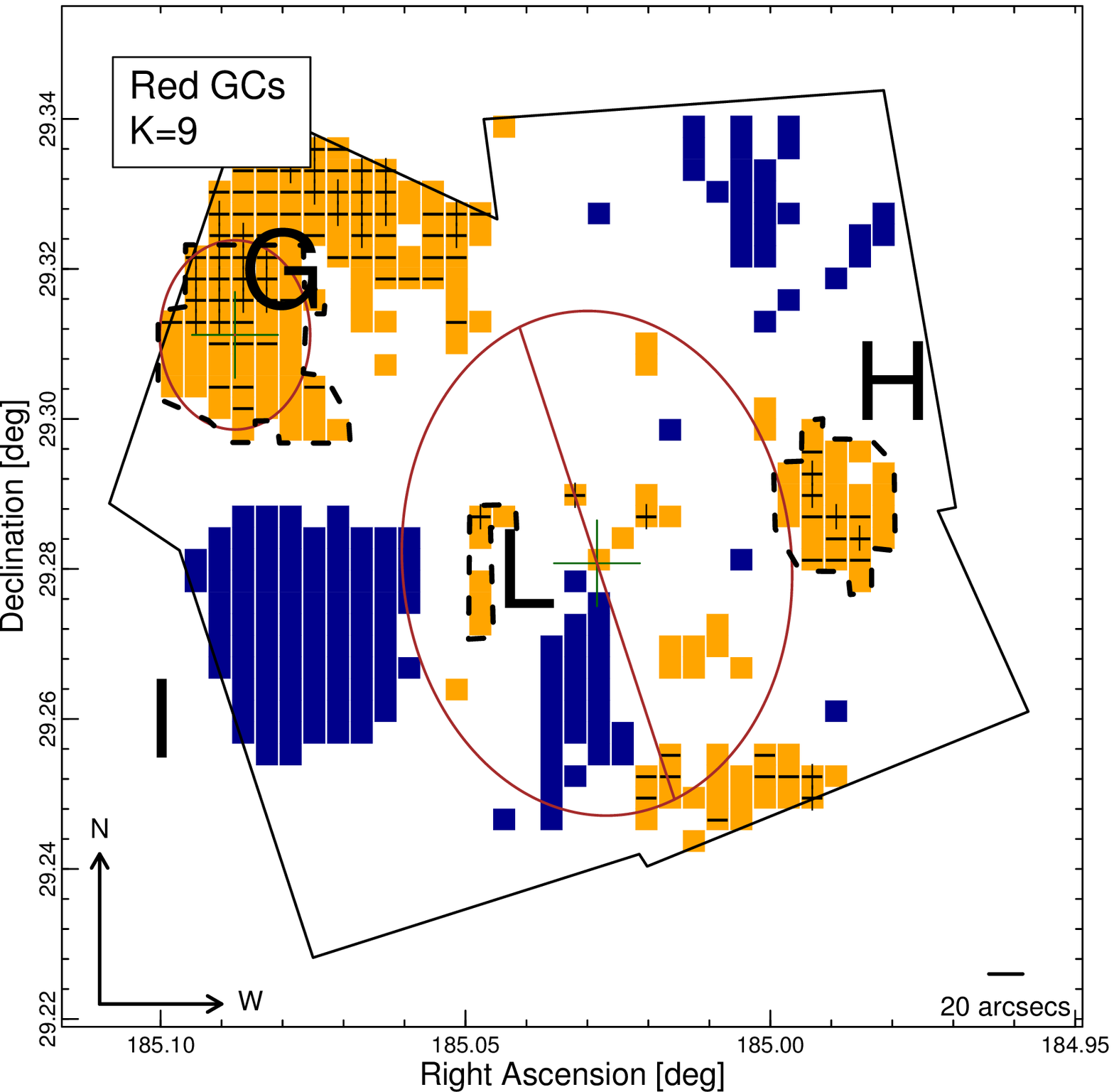}	
	\includegraphics[height=6cm,width=6cm,angle=0]{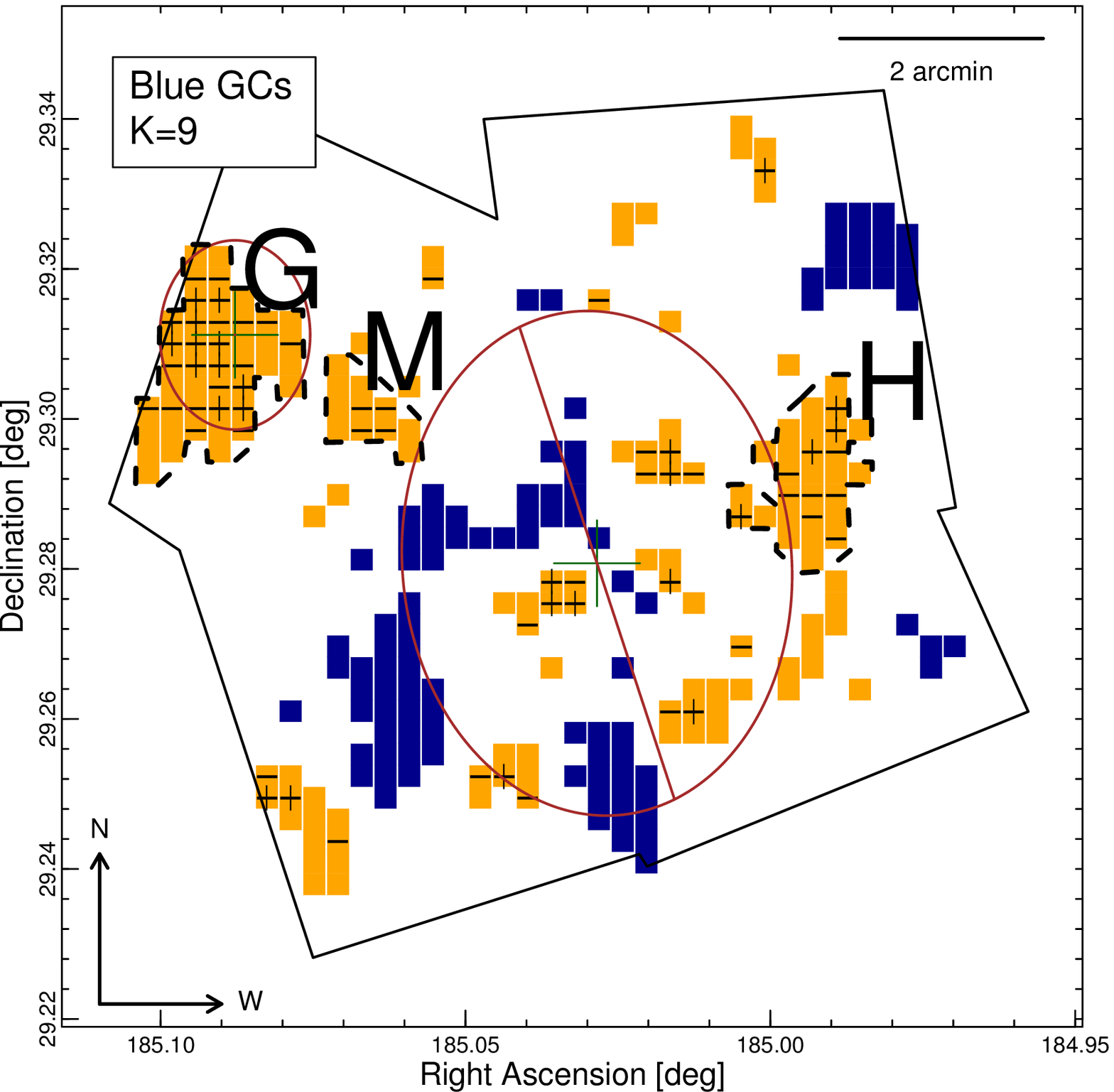}\\
	\caption{Upper panels: from left to right, residual maps of all, red and blue GCs of 
	NGC4278 from the U13 catalog, for $K\!=\!7$, $K\!=\!9$ and $K\!=\!9$ respectively. 
	Pixels are color-coded according to the number of $\sigma$ each pixel deviates from the average of the
	simulated density. Darker colors indicate larger
	residuals: blue, negative; orange, positive. The small $+, -$ and $=$ signs within each 
	pixel indicate positive, negative or null residuals respectively. The footprint of the HST observations
	used to extract the catalog of GCs and the $D_{25}$ ellipses of NGC4278 and NGC4283 are also shown. 
	Lower panels: same as above, with only pixels associated to positive and negative residuals larger than 
	1$\sigma$ shown.}
	\label{fig:2dmapsNGC4278_gc_new}
\end{figure*}

Using U13, we produced density and residual maps of the entire sample of GCs and of 
the two color classes obtained with the 
threshold $g\!-\!z\!=\!1.078$ for values of $K$ ranging from 2 to 10 
(see Section~\ref{sec:results} for more details) and with $N\!=\!5000$ simulations. The 
probability of a simulated
map showing the same number of ``extreme'' pixels of the observed density map becomes negligible 
for $K\!=\!7$ and $K\!=\!9$ for 
all and red/blue GCs, respectively. The residual maps for all, red and blue U13 GCs 
are shown in Figure~\ref{fig:2dmapsNGC4278_gc_new} (upper panels). 
The most significant features of the residual maps (clearly shown in the maps of the $>1\sigma$ significance
pixels - mid panel in Figure~\ref{fig:2dmapsNGC4278_gc_new}) are the large
over-density associated to the region occupied by the nearby galaxy NGC4283 in the N-E corner of the observed 
field (``G''), the 
significant over-density along the $D_{25}$ on the W side of NGC4278 (``H''), and the large under-density located on the 
E side of the NGC4278 galaxy, south of the NGC4283 galaxy (``I''). All of these structures are clearly 
visible in the residual maps generated either using the entire GC sample or the two colors sub-samples, 
with significance larger than 8$\sigma$ in all three cases. A peculiar but less significant structure 
($<1.5\sigma$ for all GCs, 
$\sim2.7\sigma$ for red GCs), located within the $D_{25}$ isophote of the 
NGC4278 galaxy and west of the center of the galaxy (``L''), is visibile in the residual maps produced using 
all GCs and red GCs. This elongated structure is similar to the ``streamer'' (``B'') identified in the residual map 
of the sample
of LMXBs from the catalog presented by~B09 (see Figure~\ref{fig:2dmapsNGC4278_lmxb}). 
While this structure is not significant {\it per se}, the fact that it is close to a high-significance under-density 
makes it unlikely that it is due to a random density contrast. Another feature that seems to point towards 
the center of NGC4278 is clearly visible in the residual map generated from the entire GCs sample 
(``M'' in lower-left plot of Figure~\ref{fig:2dmapsNGC4278_gc_new}). This over-density is significant at a
$\sim4\sigma$ level in the residual map of all GCs and at a $\sim3.2\sigma$ level in the residual map
generated by blue GCs shown in Figure~\ref{fig:2dmapsNGC4278_gc_new} (lower right). 
 
\begin{figure*}[]
	\includegraphics[height=8cm,width=8cm,angle=0]{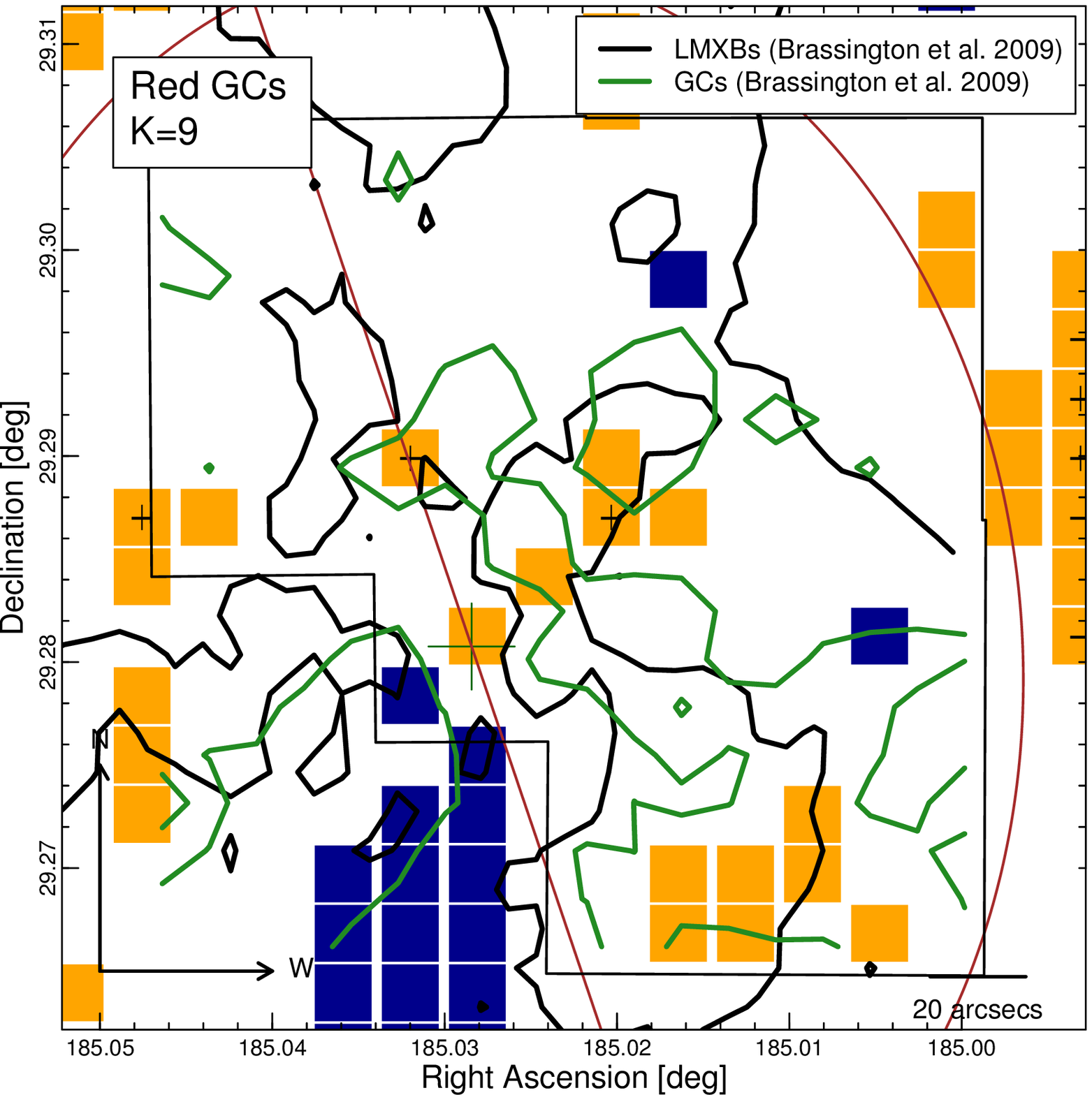}
	\includegraphics[height=8cm,width=8cm,angle=0]{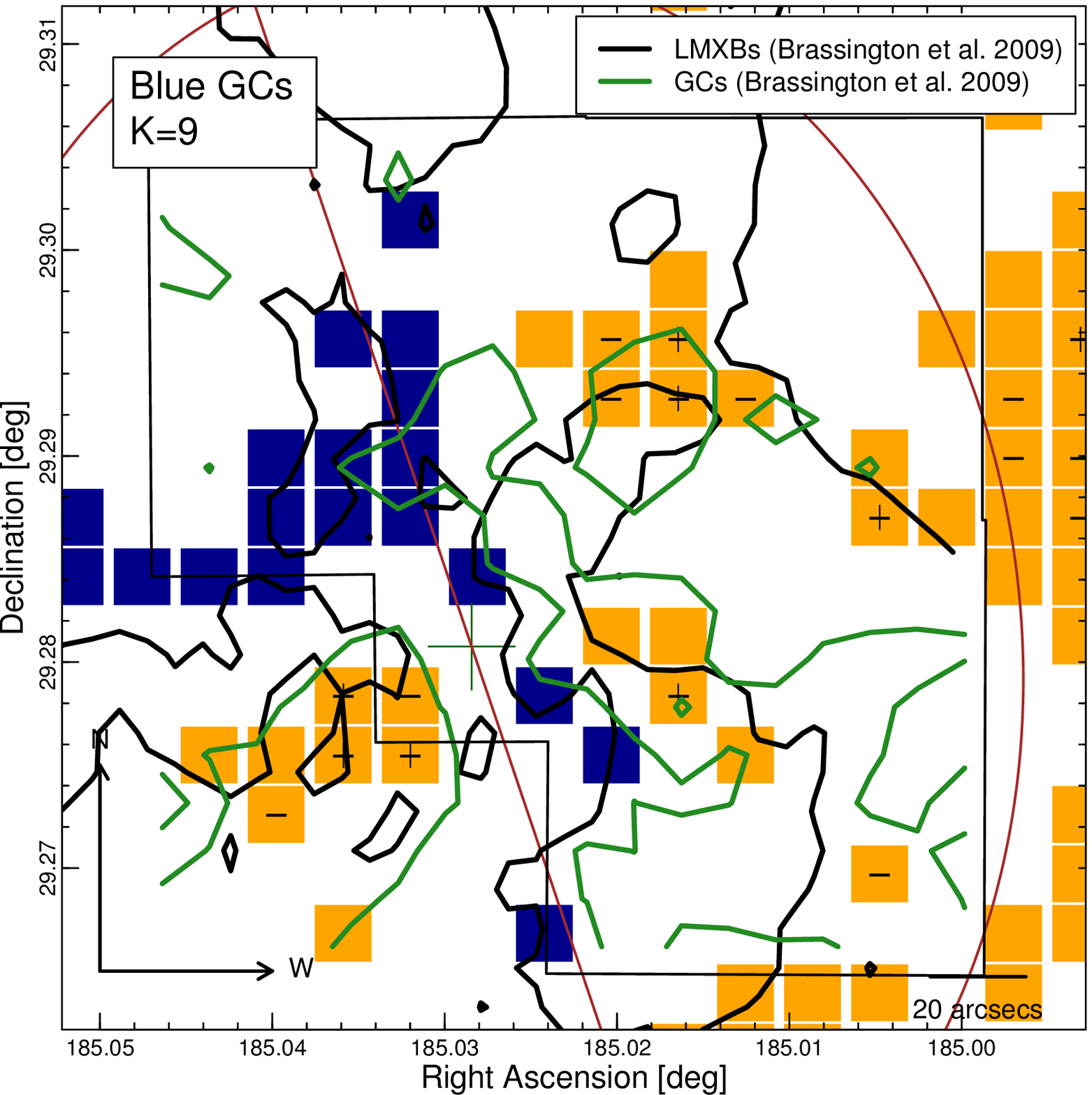}
	\caption{Residual maps for red GCs (left) and blue GCs (right) of the U13 catalog 
	for $K\!=\!9$, zoomed into the region observed by~\cite{kundu2001} 
	and used to extract the catalog of GCs used in~B09. The $D_{25}$ isophote of 
	NGC4278 and the major axis from are shown. 
	The thick green and blue lines show the over-densities in the residual maps derived 
	from the distributions of catalog of LMXBs and GCs discussed in~B09.}
	\label{fig:2dmapsNGC4278_gc_new_inset}
\end{figure*}

Figure~\ref{fig:2dmapsNGC4278_gc_new_inset} shows the position of the pixels associated to 
$\geq1\sigma$ residuals for the red GCs (left) and the blue GCs (right) for $K\!=\!9$ in the region of 
the field covering the region of NGC4278 covered by the samples of GCs and LMXBs from~B09. 
The solid black and green lines represent the contours of the over-density in the residual maps of 
the LMXBs and GCs of the~B09 paper. Even though the $>1\sigma$ pixels in the significance maps 
produced with U13 GCs are not as spatially clustered as in 
the maps generated from the~B09 GCs, 
they are mostly located within the outlines of the structures
from the residuals maps of LMXBs and GCs from~B09 (compare with 
pixels occupied by over-densities in Figure~\ref{fig:2dmapsNGC4278_lmxb} and 
Figure~\ref{fig:2dmapsNGC4278_gc} for LMXBs and GCs respectively). In more detail, 
a group of positive residual pixels in the red GCs map are located in the external region 
of the arc feature (S-W of the center of NGC4278). Another less significant structure is approximately
placed along the N-E streamer. In the residual maps generated with the blue GCs, a fairly significant
over-density ($\sim3\sigma$) is located on the N tip of the streamer feature, while few positive pixels
overlap the innermost section of the S-W arc.

\section{Discussion}
\label{sec:discussion}	

Our analysis demonstrates the presence of significant inhomogeneity in the projected 
spatial distribution of LMXBs in NGC4278, including an arc to the south of the 
nucleus, a radial streamer, and an over-density at large radii. These features are 
reminiscent of those reported in the elliptical galaxies 
NGC4261~\citep{zezas2003} and NGC4649~\citep{dabrusco2013b}. 
The study of the 2D distribution of LMXBs in NGC4261~\citep{zezas2003} was based on 
a relatively short single Chandra observation, leading to 
the detection of a small number of highly luminous sources. These results may be 
questioned by following deeper studies because of source variability (A. Zezas, work 
in preparation). In the case of both NGC4278 and NGC4649 the lists of LMXBs were 
derived from the analysis of several coadded deep observations taken over varied 
time spans, which would average over variability effects. Moreover, in NGC4278 
a later snapshot, not included in~B09, was obtained to study 
long-term variability of luminous LMXBs~\citep{brassington2012}. The full data set 
confirms the spatial features discussed in this paper (see Figure 1 of~\cite{pellegrini2012}).

The arc and the over-density at large radii are seen only in the 2D distribution of LMXBs 
associated with GCs. The 2D spatial distributions of red and blue GCs also show 
significant features. 
In particular, within the central region of the galaxy, an over-density in the red 
GC distribution is coincident with the LMXB arc, and features can also be seen 
loosely correlated with the LMXB streamer. At larger radii, apart from a significant 
over-density of GCs associated with the position of the galaxy NGC4283, two main
features dominate the spatial distribution of GCs: a significant under-density region
located E of NGC4278 and close to the $D_{25}$ isophote of NGC4283, most evident 
in the red GCs map, and a significant over-density on the W side of NGC4278, which
is clearly visible in all residual maps. 

As discussed in our previous work~\citep{zezas2003,bonfini2012,dabrusco2013a,dabrusco2013b}, 
these spatial features are strongly suggestive of the effects of galaxy growth by accretion of 
satellites and merging with neighboring galaxies. These phenomena have been observed in our 
Local Group, and are connected with streams of GCs~\cite[e.g.,][]{belokurov2006,ibata1994,mcconnachie2009}. 
GCs are likely to be the individually detectable fossil remnants of these 
events in more distant galaxies~\citep[see e.g., the simulation of][]{penarrubia2009}. 
Similarly, X-ray binaries are also individually detectable tracers of their parent 
stellar population~\citep{fabbiano2006}.

The three elliptical galaxies we have investigated so far, present a range of 
properties. NGC4261 and NGC4649 are giant ellipticals with $\log{L_{K}}\!=\!11.4\ L_{K,\sun}$, 
while NGC4278 is of intermediate stellar mass, with 
$\log{L_{K}}\!=\!10.8\ L_{K,\sun}$~\citep{devaucouleurs1991}; all of them have 
old stellar populations, with ages ranging from 12 to 16.3 
Gy~\citep[see compilation in][]{boroson2011}. While NGC4649 is in Virgo, 
NGC4261 and NGC4278 are in small groups~\citep[see][]{garcia1993}. NGC4261 
hosts a luminous radio AGN~\citep[3C270,][]{birkinshaw1985}, NGC4649 hosts a 
low-luminosity nuclear radio source~\citep{dressel1985} and NGC4278 has a 
LINER nucleus, which is also detected in the radio~\citep{younes2010,terashima2003}. 
Of the three galaxies, NGC4649 is the only one associated with a 
prominent luminous extended hot halo~\citep[e.g.,][]{humphrey2013}. 

These three galaxies all present some sign of merging 
or accretion of external matter, other than our results: NGC4261 has clear boxy 
isophotes~\citep{nieto1989}, usually suggestive of merging~\citep{kormendy1996}. 
Recent kinematical measurements~\citep{arnold2013} 
suggest that NGC4649 might be a major dry merger remnant, showing disk-like 
outer rotation, which suggests a massive lenticular galaxy (S0)
progenitor. NGC4649 has a near neighbor, the spiral galaxy NGC4647, with which 
it is likely to be tidally interacting~\citep{degrijs2006}. Both NGC4261 and NGC4649
are early-type SuperNova (SN) Ia host galaxies. SN 1939B, located close to the outer 
side of the large-scale arc structure of GCs discussed by~\citep{dabrusco2013a} in 
the S-W quadrant of NGC4261, was discovered by~\cite{zwicky1939} and was later
classified as type Ia SN. Type Ia SN 2004W~\citep{moore2004} was located 
51.6$^{\prime\prime}$ W and 78.7$^{\prime\prime}$ S of the nucleus of NGC4649, 
where a low significance over-density of GCs has been determined (see right plot in 
Figure~3 of~\cite{dabrusco2013b}). The presence of type Ia SNs in these two galaxies
hints at intermediate age stellar population that could be linked to past merger and/or 
interaction events. NGC4278 is relatively more isolated, although 
also has a near neighbor, NGC4283. Moreover, it has 
a complex multi-phase ISM~\citep[see][]{pellegrini2012}.

NGC4278 has a large massive HI ring, extending for 37 kpc~\citep{raimond1981}, 
which may be slowly accreting~\citep{morganti2006}. This disk extends down to 
$\sim$1 kpc, where a ionized rotating disk is found~\citep{sarzi2006}.~\cite{morganti2006} 
show that the HI ring in NGC4278 has a complex structure, with two faint tail-like structures 
at large radii, pointing towards S-W and N-E of the center of the galaxy respectively (see Figure~1
in~\cite{morganti2006}). While both HI
tails are more spatially extended than the over-densities in the GCs and LMXBs 
distributions discussed in this paper, we notice that the GCs over-density H (lower panels of 
Figure~\ref{fig:2dmapsNGC4278_gc_new}) overlaps
the base on the S-W HI tail-like structure.~\cite{morganti2006} discuss the origin of 
the neutral gas in early-type galaxies, arguing that while merger
or accretion events are thought to be the origin of some of the structures observed in 
the spatial distribution of HI, the lack 
of perturbation in the stellar population of some galaxies suggests that cold accretion 
from the Inter-Galactic Matter (IGM) could also be responsible for the presence of 
neutral gas. The HI disk in NGC4278 is co-located 
with a disk of X-ray emission and with complex and irregular dusty filaments, which 
may be streaming towards the nucleus~\citep{pellegrini2011,lauer2005,tang2011}. 
These features have also been interpreted as due to slow accretion of 
the HI gas, unaccompanied by star formation, given the old stellar population age; 
the X-ray disk may be the result of conduction heating of the incoming material by 
a more diffuse hotter X-ray emitting ISM~\citep[][and references therein]{pellegrini2011}. 
However, the inhomogeneity we have discovered tells a different, perhaps 
complementary, story. Given the old stellar age and the regular appearance of the 
stellar spheroid, any LMXBs originating for the evolution of native binary systems 
in the stellar field, ought to be distributed smoothly, closely following 
the stellar surface brightness. If the GC populations themselves were also formed 
early in the history of NGC4278, the same would hold for the distribution of GCs 
and their associated LMXBs. While this is true overall in the radial distribution of 
LMXB number density~(see Figure 4 of B09), our results 
clearly show that there are local perturbations to this picture. Therefore some of the accretion 
events in NGC4278 may have involved the disruption of dwarf companions or 
perhaps result in some localized star formation, that while not visible against 
the preponderant old stellar population may be revealed by the LMXB and GC 
inhomogeneities.

\section{Summary and Conclusions}
\label{sec:conclusions}

We have discovered statistically significant spatial inhomogeneities in the 2D 
distributions of LMXBs and GCs in the intermediate mass elliptical NGC4278. In 
particular, in the inner region of the galaxy we detect an arc-like feature in the 
LMXB distribution, which is also 
reflected in a local over-density of the 2D distribution of red GCs. At larger radii, 
the distribution of GCs is dominated by a very significant over-density
likely associated to the nearby galaxy NGC4283. Two other significant density 
structures are visible outside of the $D_{25}$ isophote to the 
W and E of the center of NGC4278, associated to an over-density and an 
under-density respectively. 

These features are 
suggestive of streamers from disrupted and accreted dwarf companions, of which 
LMXBs and GCs may be the fossil remnants. NGC4278 is the third galaxy where 
we have detected this type of feature. Similarly for two previous cases, 
NGC4261~\citep{dabrusco2013a} and NGC4649~\citep{dabrusco2013b}, the 
availability of observations has provided us with the ability to characterize the 
whole GC population of the galaxy and thereby identify the large-scale disturbances
with the system. Our exploratory work begs the question of how frequent these features may be. 
If they are common, they may 
revealed widespread continuing galaxy evolution by accretion, in agreement with 
the predictions of the $\Lambda$CDM theory~\citep{dimatteo2005}. We have 
demonstrated that the study 
of the 2D distributions of GCs systems (and LMXBs, when available) is a valuable 
tool in these investigations. The next step will be to obtain good systematic optical 
coverage of a well chosen sample of elliptical galaxies.

\acknowledgements

We thank J. Strader and A. Zezas for useful discussions and comments that have helped 
to improve the paper. This work 
was partially supported by the {\it Chandra} X-ray Center (CXC), which is operated by the 
Smithsonian Astrophysical Observatory (SAO) under NASA contract NAS8-03060.

{}

\end{document}